\documentclass[twocolumn,tight,times]{aastex63}

\usepackage{amsmath,amstext}
\usepackage[figure,figure*]{hypcap}
\usepackage{newtxmath} 
\usepackage{longtable}

\newcommand{\mteff}{\tau_\mathrm{eff}}
\newcommand{\teff}{$\mteff$}
\newcommand{\lya}{Ly$\alpha$}
\newcommand{\kms}{km~s$^{-1}$}
\newcommand{\zem}{$\mathrm{z_{med}}$}
\newcommand{\hiz}{$\mathrm{z_{high}}$}
\newcommand{\loz}{$\mathrm{z_{low}}$}

\newcommand{\Aval}{0.159}  
\newcommand{\Aerr}{0.001}  
\newcommand{\Bval}{-2.022}  
\newcommand{\Berr}{11.60}  

\begin{document}

\title{Effective Opacity of the Intergalactic Medium from Galaxy Spectra Analysis}

\correspondingauthor{Jose Monzon}
\email{jsmonzon@ucsc.edu}

\author[0000-0002-9986-4604]{Jose S. Monzon}
\affiliation{University of California, Santa Cruz; 1156 High St., Santa Cruz, CA 95064, USA}

\author[0000-0002-7738-6875]{J. Xavier Prochaska}
\affiliation{University of California, Santa Cruz; 1156 High St., Santa Cruz, CA 95064, USA}

\author[0000-0001-9299-5719]{Khee-Gan Lee}
\affiliation{Kavli Institute for the Physics and Mathematics of the Universe (WPI), University of Tokyo, Kashiwa 277-8583, Japan}

\author[0000-0002-0302-2577]{John Chisholm}
\altaffiliation{Hubble Fellow}
\affiliation{University of California, Santa Cruz; 1156 High St., Santa Cruz, CA 95064, USA}

\begin{abstract}

We measure the effective opacity (\teff) of the Intergalactic Medium (IGM) from the composite spectra of 281 Lyman-Break Galaxies (LBGs) in the redshift range $2 \lesssim z \lesssim 3$. Our spectra are taken from the COSMOS Lyman-Alpha Mapping And Tomographic Observations (CLAMATO) survey derived from the Low Resolution Imaging Spectrometer (LRIS) on the W.M. Keck I telescope. We generate composite spectra in two redshift intervals and fit them with spectral energy distribution (SED) models composed of simple stellar populations. Extrapolating these SED models into the \lya\ forest, we measure the effective \lya\ opacity (\teff) in the $2.02 \leq z \leq 2.44$ range. At $z = 2.22$, we estimate $\mteff = \Aval \pm \Aerr$ from a power-law fit to the data. These measurements are consistent with estimates from quasar analyses at $z<2.5$ indicating that the systematic errors associated with normalizing quasar continua are not substantial. We provide a Gaussian Processes model of our results and previous \teff\ measurements that describes the steep redshift evolution in \teff\ from $z = 1.5 - 4$.

\end{abstract}


\section{Introduction}
\label{sec:intro}

The Intergalactic Medium (IGM) is a diffuse gas, mainly consisting of ionized hydrogen and helium, that permeates the space between galaxies in the large-scale cosmic web. The gas is highly ionized by the extra-galactic ultraviolet background (EUVB) radiation field and takes the form of a diffuse, $T\sim10^4$K plasma. The trace fraction of hydrogen gas that remains neutral ($\chi_\mathrm{HI}$), is responsible for attenuating the radiation from the EUVB and producing the \lya\ forest \citep[see][for a review]{McQuinn_2016}. Studies on the \lya\ forest have meshed well with cosmological theory as it is the IGM, and not the galaxies it surrounds, that governs the large-scale structure of the universe. It is considered one of the most powerful cosmological probes at $z \geq 2$ as it holds the majority of baryons at all epochs \citep[e.g.][]{Becker_2007} 

\cite{Gunn_Peterson_1965} were the first to discern that a universe filled with neutral hydrogen (HI) would be opaque in the far-UV, especially at higher redshifts which is densest. Analyzing the spectrum of any distant, rest-frame UV-emitting object, directly points to a fluctuating and photo-ionized gas that at a given redshift, varies considerably from sight-line to sight-line \citep{Shapley_2003}. A series of studies \citep[e.g.][]{Dall'Aglio_2008, Faucher-Giguere_2008, Becker_2013} have since carried out careful measurements of how HI evolves to place statistical constraints on properties like density, temperature and composition. A solid understanding of the physical state of the IGM allows for subsequent research investigating galaxy formation \citep{Hassan_2020}, the ionization history \citep{Theuns_2002, Bernardi_2003, Kirkman_2005} and ultimately the constraints on our leading cosmological theories \citep{Rauch_1998, Becker_2007}.

Studies of the physical properties of the IGM have primarily come from the analysis of the mean optical depth of HI ($\tau$) observed in the spectra of distant Quasi-Stellar Objects \citep[QSOs;][]{Prochaska_2009, Becker_2007, Faucher-Giguere_2008, Kirkman_2005}. QSO's peak in the UV because of their hot accretion disks \citep[AGN;][]{Meiksin_2009}. As a QSO's radiation traverses the space between galaxies, a series of absorption lines populate the rest-frame spectrum blueward of 1215\AA. Because the IGM is inhomogeneous, photons interact with the intervening gas at different redshifts, causing absorption features across a multitude of wavelengths; the so-called \lya\ forest. For sufficiently distant objects ($z > 5$), where the IGM is the densest \citep{Mcdonald_2006}, the absorption lines become so numerous that more than 70\% of the flux is absorbed in the \lya\ forest ($\sim$ 1020-1210\AA). 

One can directly measure values describing the attenuation from the spectra of distant objects by estimating the underlying continuum, a process that becomes increasingly difficult at higher redshifts \citep[e..g][]{Kirkman_2005}. QSOs are much brighter and therefore easier to observe at higher redshifts, but the \lya\ forest can be observed in the spectra of any distant, UV-emitting source. In fact, the $z>4$ EUVB is thought to be dominated by a population of faint UV-emitting galaxies in addition to bright QSOs at $g \sim 23$ magnitudes \citep{Lee_2014}. Their contribution to the EUVB is caused by the young and massive stars they harbor. We set out to measure the effective \lya\ opacity of the IGM, \teff\, using the spectra of these Lyman-Break galaxies (LBGs) by exploiting their high number density \cite[hereafter L18]{Lee_2018}.

LBGs are star forming galaxies whose emission peak in the rest-frame UV and are selected based on their emission blueward of \lya\ in a given filter set \citep{Steidel_1996}. The original term 'LBG' describes star-forming galaxies selected at $z \geq 3$ by their IGM absorption, but the CLAMATO team use the term to cover all $z \geq 2$ galaxies with a far-UV continuum. The stacked spectra of LBGs have been used to constrain the dust attenuation curve \citep{Reddy_2016} and investigate spectral features attributable to hot stars, HII regions and outflowing gas \citep{Shapley_2003}. \cite{Thomas_2017} analyzed galaxy spectra to estimate \teff\ at $2.5 < z < 5.5$.  Using only LBG spectra, they provided an assessment of \teff\ without the systematic errors associated with normalizing quasar spectra.

In this work, we leverage the blue sensitivity of the Keck/LRIS spectrograph to measure \teff\ from low S/N LBGs spectra  at $z \lesssim 2.5$.  Similarly, we set out to test previous work analyzing the effective opacity of the IGM by generating an estimate independent of the challenges associated with normalizing quasar spectra. Crucially, $z < 2.5$ is the regime where quasar measurements have traditionally been anchored on the grounds that one can more accurately estimate quasar continua at lower opacity.

In the following sections of this manuscript we: [2] present the CLAMATO data sample [3] describe our methodologies for creating composite spectra and fitting SED models, [4] report our measurements of \teff and compare to previous studies, and [5] summarize and discuss this work’s findings. Throughout the paper, we adopt a concordance Lambda Cold-Dark-Matter ($\Lambda$-CDM) cosmology with $\Omega_{\Lambda}$ = 0.7, $\Omega_{m}$ = 0.3 and h = 0.7.

\section{The CLAMATO Observations and Sample Selection}
\subsection{CLAMATO}
\label{subsec:clamato}

Our sample of galaxies was drawn from the 2016 and 2017 releases of the COSMOS Lyman-Alpha Mapping And Tomographic Observations (CLAMATO) which were measured by the Low Resolution Imaging Spectrometer (LRIS) on W.M. Keck I telescope \citep{LRIS}. CLAMATO began operations in 2014 with the main goal of mapping the \lya\ forest tomography of the foreground IGM (these pilot observations were not applicable to our analysis). \cite{Lee_2014} found that because galaxies dominate the foreground UV luminosity function at faint magnitudes \textit{g} $\sim$ 23 \citep{Reddy_2008}, LBG spectra would almost exclusively compose the 3D tomographic reconstruction.

CLAMATO is designed to systematically observe faint ($23 \lesssim \textit{g} \lesssim 25$) UV-emiting sources from $2 < z < 3$, at high area densities ($\sim 1000 deg^{2}$) and L18 reports using a total of 240 background galaxies and QSOs within a 0.157 square degree section of the COSMOS field. They also report estimated redshift values and spectra on an additional 437 objects for a total 677 reduced sources. The COSMOS field \citep{Scoville_2007} is in the Northern Hemisphere and spans 2 square degrees. It offers a large selection of $g$-band star forming galaxies, covers a significant scale in the transverse direction ($\sim 10$\,Mpc) and has measurements of redshifts for the objects in the survey. 

The target selection procedure for CLAMATO depends on the magnitude and probability of success, initial prioritization based on redshift, and the subsequent slit mask designs. As the COSMOS field has a rich selection of spectroscopic and multi-wavelength imaging data, L18 built CLAMATO from existing redshift catalogs \cite{Lilly_2007, LeFevre_2015, Kriek_2015, Nanayakkara_2016} that covered their desired wavelength range ($3700\AA < \lambda < 4300\AA$). To select targets, L18 fed the combined spectroscopic and photometric catalogs to an algorithm which prioritizes background $g$-band sources in the redshift range of $2.25 \lesssim z \lesssim 2.45$. The algorithm prefers brighter sources due to slit-packing constraints but selected targets as faint as $g = 25.3$.

\begin{figure*}[ht]
    \begin{center}
    \includegraphics[scale=.5]{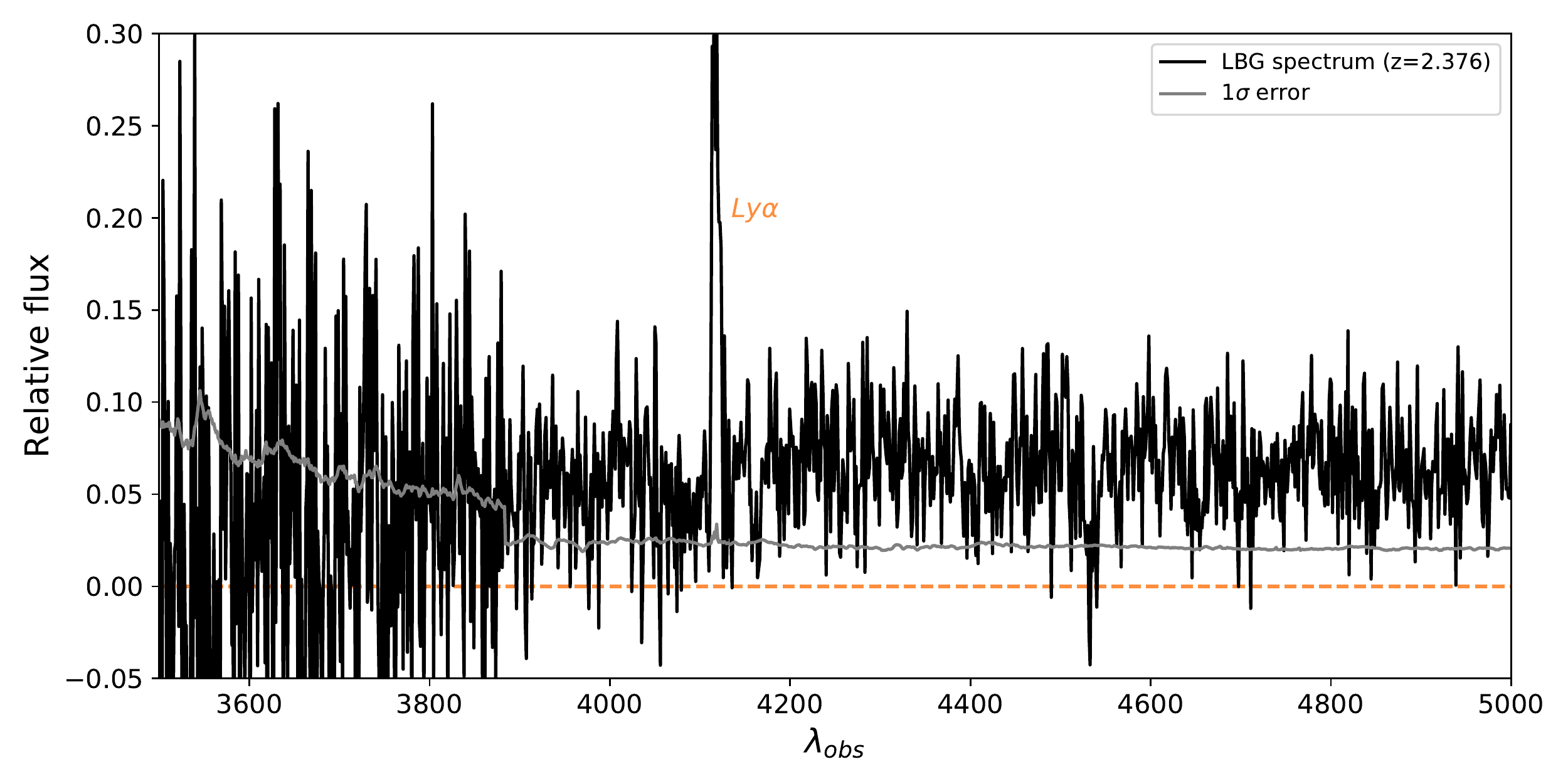}
    \caption{An example spectrum of a Lyman Break Galaxy from the CLAMATO data release that fits our sample selection criteria. The spectrum shows a bright \lya\ emission feature and weaker ISM features that are difficult to distinguish from the noise. An error spectrum (shown in grey) is reported with each source in the CLAMATO release.}
    \label{fig:exspec}
    \end{center}
\end{figure*}

Observations for CLAMATO lasted a total of 15.5 nights of which about 60 hrs were spent on sky with typical total exposure time per object lasting $\sim$ 9000s. LRIS was configured with the 600/4000 grism to achieve an approximate resolution R $\equiv \lambda/\Delta\lambda \approx 1000$ with 1$"$ slits between the observer-frame wavelengths of 3700A and 4400A on its Blue channel. As expected with such faint and distant sources and an average seeing of 0.7$"$, the spectra have low signal-to-noise, S/N $< 3$ per \AA. The data were then processed using the LowRedux routines from the XIDL software package\footnote{http://www.ucolick.org/$\sim$xavier/LowRedux}. Figure \ref{fig:exspec} is an example of a reduced galaxy spectrum taken from the CLAMATO release described in L18. 

L18 then assigned confidence ratings from $0-4$ when estimating redshifts for each source, 0 being no attempt at all (normally reserved for corrupted data) and 4 being high confidence based on multiple lines. L18 reports that 66\% of the objects in the sample had confidence ratings $\geq 3$. The majority of less secure redshifts are for low priority sources used to fill spare slit space that often yielded spectra too noisy to identify. Approximately 95\% of the objects with confidence ratings $\geq 3$, were identified as galaxies using LBG templates from \cite{Shapley_2003}, while the other 5\%, were distinguished as broad-line quasars. For a more detailed outline of the selection algorithm, instrument specifications and preliminary data reduction please see L18.

\subsection{Sample Selection}
\label{subsec:sample}

Measuring the opacity, $\tau$, from an individual object yields a single realization of the stochastic IGM. The \textit{effective} opacity of the IGM \teff\ is the average estimated over many sight-lines. We measure \teff\ using a composite or ``stacked" spectrum, which is essentially an average of flux values, at each wavelength. Alternatively, we could have measured the opacity from several different spectra, and then averaged, to yield \teff. There are two main justifications for why we chose to average our data before measuring the opacity. First, for small redshift variations, the observed continua of LBGs (or QSOs) are consistent across sight-lines, so a composite spectrum can be modeled by a single SED. Second, and more importantly, stacking improves the S/N allowing us to more accurately model the SED redward of \lya.

\begin{figure}[ht]
    \begin{center}
    \includegraphics[width=\columnwidth]{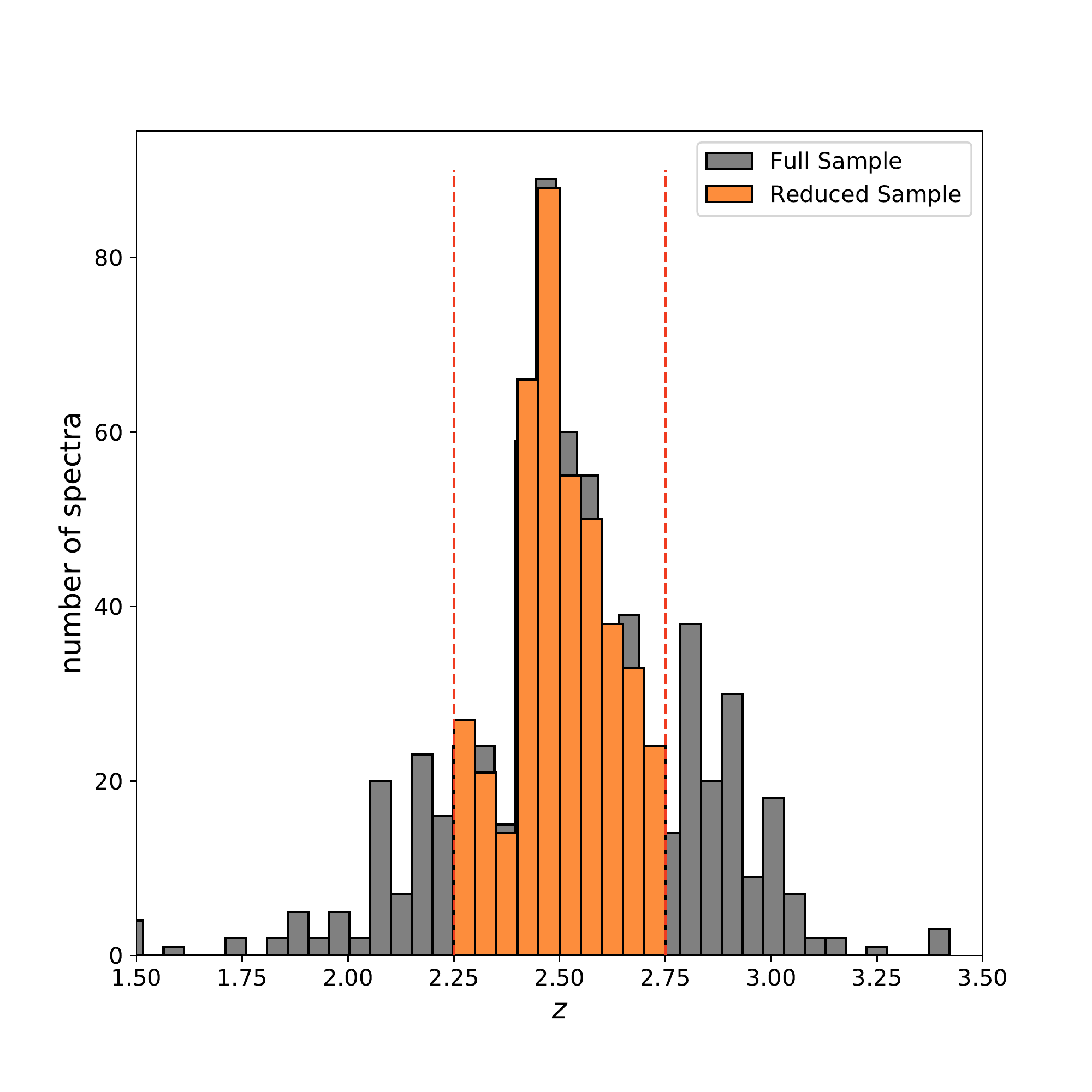}
    \caption{Redshift distribution of the complete CLAMATO sample (show in grey) vs.\ the reduced sample satisfying our selection criteria (show in orange).}
    \label{fig:clamatohist}
    \end{center}
\end{figure}

To account for the fact that we are sampling the IGM with sight-lines corresponding to objects that are not at identical redshifts, we organize the CLAMATO spectra into small redshift bins of $\Delta z = 0.25$. This yields a median redshift \zem, which serves as a reference for the \teff\ values from the \lya\ forest. In total, there are 566 CLAMATO galaxy spectra in the $2.0 \leq z \leq 3.0$ interval with the majority of these sources between $2.25 < z < 2.75$ (see figure \ref{fig:clamatohist}). We only used the $2.25 < z < 2.75$ interval because the bins to either side of it, do not contain enough spectra to create an adequate stack. We split the majority interval into two redshift intervals: \loz\ from $2.25 < z < 2.50$ and \hiz\ from $2.50 < z < 2.75$, for a combined total of 416 galaxy spectra.

\begin{figure}[ht]
    \begin{center}
    \includegraphics[width=\columnwidth]{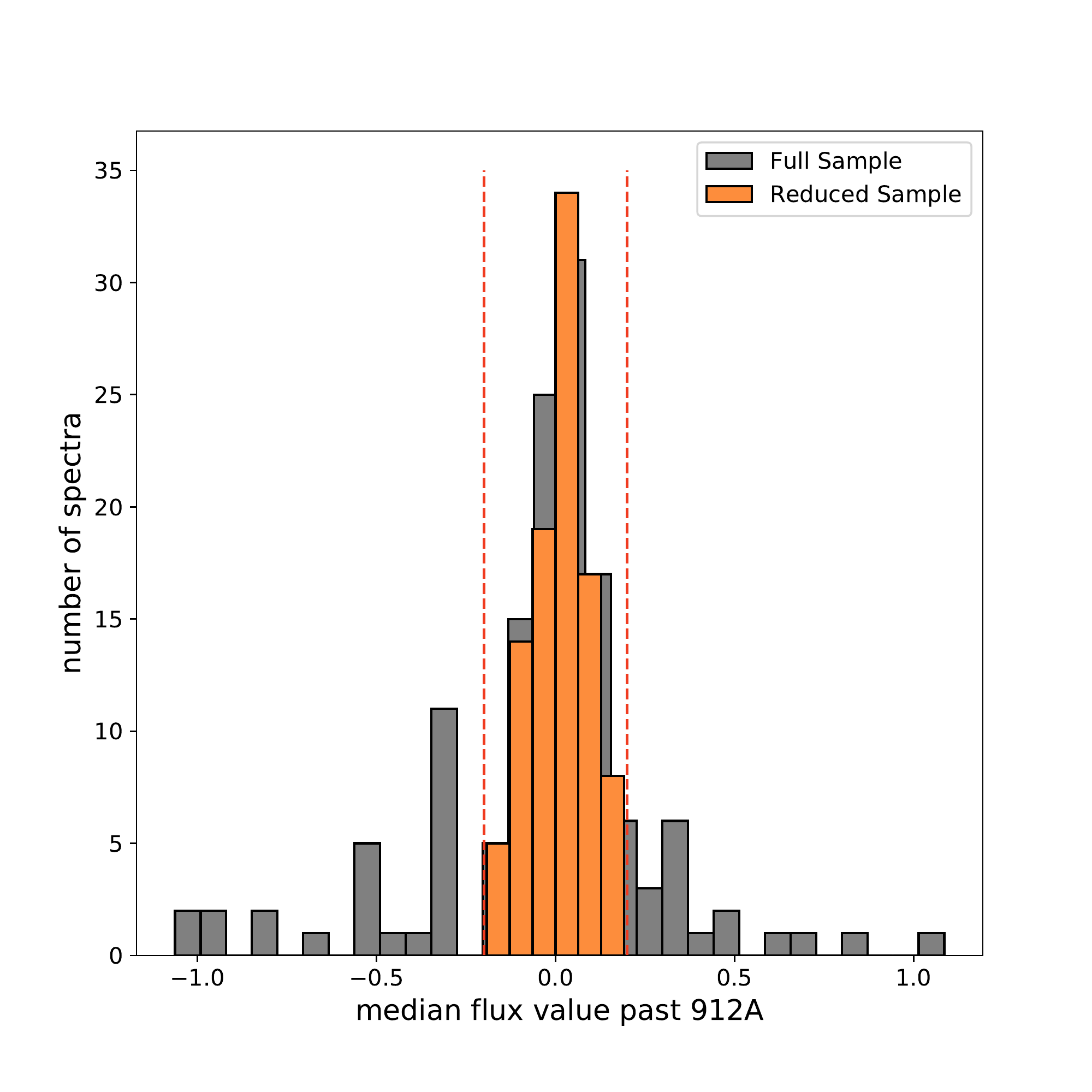}
    \caption{The complete distribution of median fl1ux values taken from spectra that extend blueward of the Lyman Limit (shown in grey). Those that are within the $\pm\ 0.2$ cut-off (show in orange) are still considered viable. We cut our sample aggressively, excluding 49 spectra, so as to not skew the continuum blueward of $\sim 1130$\AA.}
    \label{fig:ll_cut}
    \end{center}
\end{figure}

\begin{figure}[ht]
    \begin{center}
    \includegraphics[width=\columnwidth]{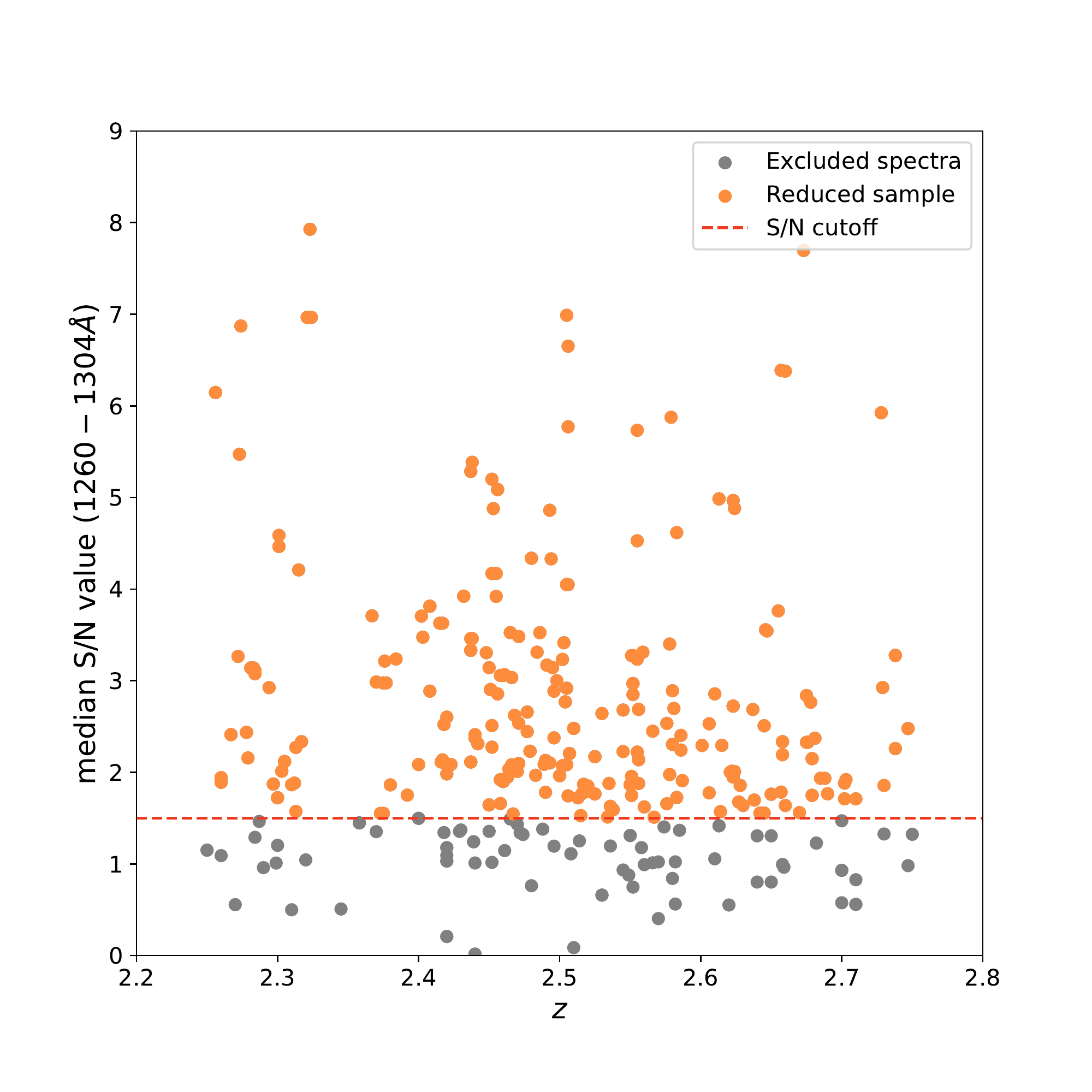}
    \caption{A scatter plot of our data's spectroscopic redshift vs the S/N calculated in the 1260-1304\AA\ range for both the \loz\ \& \hiz\ intervals. The grey objects have been excluded by S/N cut.} 
    \label{fig:noise_scatter}
    \end{center}
\end{figure}

Of our 416 galaxy spectra, several cover wavelengths blueward of the rest frame Lyman limit (912\AA). For these spectra, we measure the median flux per pixel for wavelengths below the Lyman Limit and exclude those with values outside of the $\pm 0.2$ median flux interval (see figure \ref{fig:ll_cut}). We expect these spectra to have errors in their fluxing or sky subtraction as significant signal past the Lyman Limit is highly improbable. We further cut down our sample by imposing a blanket S/N limit, using the mean flux value in the wavelength range of 1260-1304\AA. We found that a cut-off S/N = 1.5 excluded the poorest spectra without discarding the majority of the sample (see figure \ref{fig:noise_scatter}). After these two cuts, we were left with 137 in the \loz\ interval and 142 in \hiz\ interval for a combined total of 279 galaxy spectra. The \loz\ interval has a median redshift value of 2.43 and a standard deviation of 0.074. The \hiz\ interval has a median redshift value of 2.58 and a standard deviation of 0.069. See appendix table \ref{tab:clamato} for the selected sample of galaxy spectra from the CLAMATO survey.

\section{Composite Spectra}
\label{sec:composite}

\subsection{Stacking}
\label{subsec:stack}

The process that follows describes the preparation of individual LBG spectra prior to stacking:
\begin{itemize}
  \item To correct extinction from the Galactic interstellar medium (ISM), we passed each spectrum through a dereddening process based on the 3D Sky Map of \cite{Green_2018}. The Sky Map, given an object's coordinates, reports $E(B-V)$ extinction values for the Milky Way which we applied to the flux array using the the reddening curve from \cite{O'Donnell_1994} (an updated version of \citep{CCM_1989}). No other corrections were necessary as the LRIS instrument has an atmospheric dispersion correcter and a fluxing term accounting for the atmospheric extinction.
  
  \item We normalized the spectrum using values redward of \lya, where there is no absorption features due to the IGM or ISM from 1260-1304\AA\ (between two SiII lines).
  
  \item We trimmed the edges of the spectrum, only selecting flux data between $\lambda_{\rm rest} \approx$ 1050-1400\AA.
  
  \item We shifted the spectrum to the rest frame, using the redshift values measured by CLAMATO, and rebinned to a velocity dispersion of 300 \kms\ per pixel using a common starting wavelength of 1000\AA.
\end{itemize}

We then stacked the spectra by carrying out an unweighted, arithmetic mean of the flux values per wavelength. We chose not to weigh the spectra to better reduce cosmic variance in the \lya\ forest \citep{Becker_2013}. We averaged across 137 and 142 LBG spectra for the \loz\ and \hiz\ intervals (respectively). The well behaved sections of the composites (1260-1304\AA\ for example) were left with S/N values $\sim$ 30. The \lya\ forest (1070-1170\AA) however, tended towards S/N values $\sim 10$. Because in general, all individual LBG spectra edges were quite noisy (see figure \ref{fig:exspec}) our stacks remained unconstrained blueward of $\sim 1040$ and redward of $\sim 1400$. See figure \ref{fig:model} for the results of the stacking in black.

\subsection{Bootstrapping}
\label{subsec:bootstrap}

\begin{figure*}[ht]
    \centering
    \includegraphics[scale = .5]{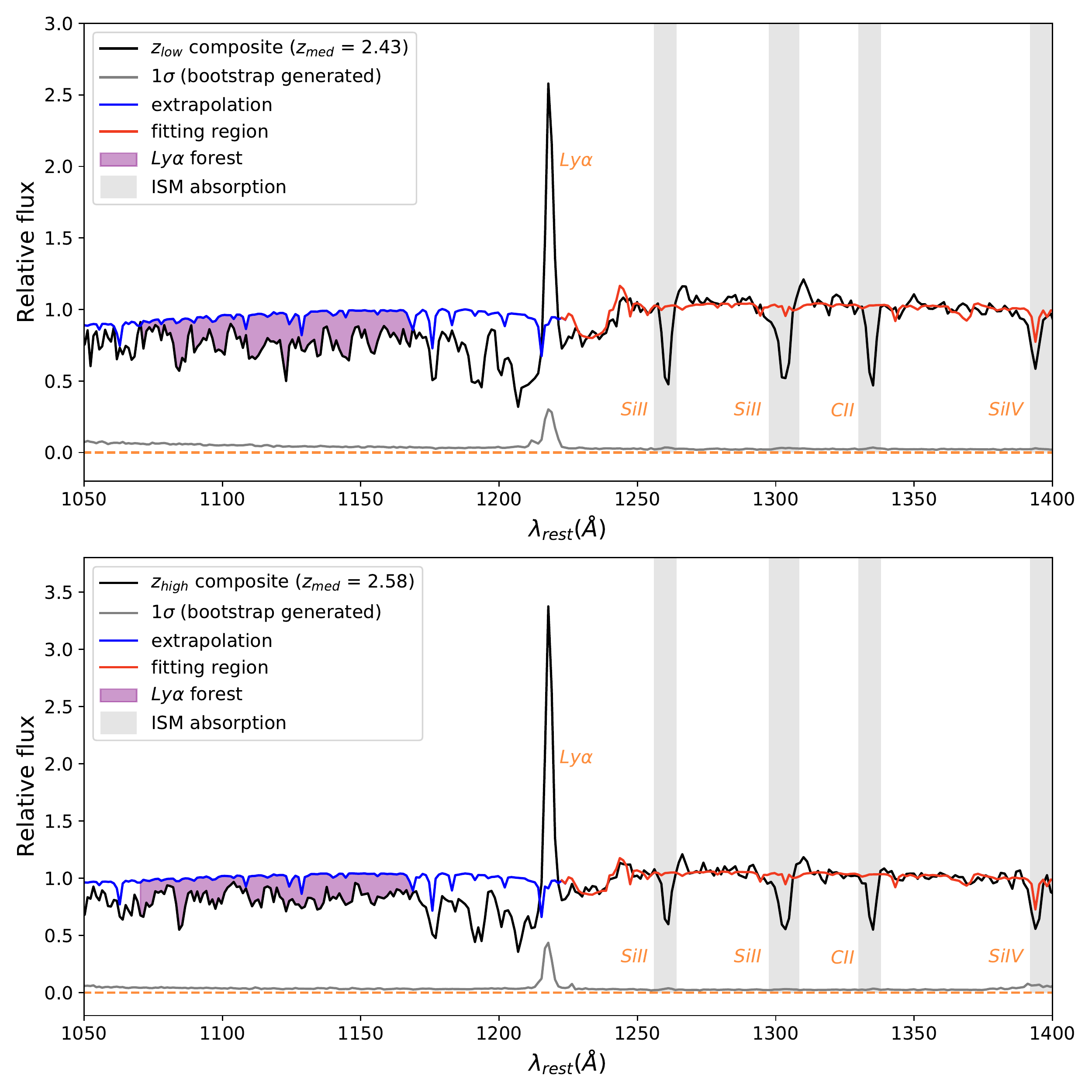}
    \caption{The un-shuffled \loz\ and \hiz\ composite spectra and their SED fits. The SED fits are two-toned, showing the region in which the fit is constrained by our data (red), and where the fit is extrapolated to measure the forest (blue). The excess flux in the model, blueward of \lya\ is caused by HI attenuation by the IGM. The error spectrum is based on the bootstrap matrix (plotted in grey; see \ref{subsec:bootstrap}). Some of the most prominent ISM transitions are denoted in grey and were not included in the fit.}
    \label{fig:model}
\end{figure*}

Our primary source of error comes from sample variance within the stack and not from the S/N values of each individual spectrum. To assess the error in \teff, we used a bootstrapping approach, following the example of \cite{Worseck_2014}. The following details our process for constructing a covariance matrix that assesses correlated errors in our \teff\ measurements in each redshift interval:
\begin{itemize}
    \item To estimate sample variance, we chose a random selection of LBG spectra, within each redshift interval (allowing for duplicates), equal to the number of spectra that comprised each original composite (137 for \loz\ and 142 for \hiz).

    \item We stacked the random selection in the same way as detailed above for creating the original composite.
    
    \item We repeated the first two steps to generate 5,000 randomized composites.
    
    \item to normalize the randomized composites, we subtracted the original composite from each of them individually.
    
    \item We compiled the randomized composites into an $I x J$ matrix (where $I$ = 5,000 and $J$ is the length of our wavelength array ($\sim 1000$)) and dotted this matrix with its transpose to create a full covariance matrix. See figure \ref{fig:model} for the the 1D diagonal results of the error analysis in grey. 
\end{itemize}

\subsection{SED Modeling}
\label{subsec:models}

To measure \teff, we estimated the unabsorbed flux of the composites in the \lya\ forest. Following the example of \cite{Paris_2011}, we extrapolated blueward of \lya\ from a well behaved section of our spectra. We modelled the unabsorbed continua using an SED modeling technique designed by \cite{Chisholm_2019} (hereafter C19) to fit simple stellar population (SSPs) models from the Starburst99 (SB99) database \citep{Leitherer_1999}. There are 50 SB99 single age, single metallicity stellar population models investigated in C19 where each model was created using  a Kroupa IMF with a high-mass exponent of 2.3, a low-mass exponent of 1.3, and a high-mass cutoff of 100~M$_\odot$. As star light between 1200-2000\AA\ is dominated by young massive O-stars \citep{Leitherer_1999}, we only investigate a narrow regime of stellar ages: 1, 2, 3, 4, 5, 8, 10, 20, 40 Myr, each with 5 different metallicities: 0.05, 0.2, 0.4, 1.0, 2.0 $Z_\odot$. The FUV stellar continuum does not dramatically change for B-star dominated stellar populations between 40-200 Myr \citep{de_Mello_2000, Rix_2004}, thus we use an upper age of 40 Myr. Each model is fully theoretical and does not include ISM lines. They were created by sampling the high-mass portion of the Hertzsprung-Russell diagram up to temperatures of 20,000 K and a high-mass cut-off of 100 $M_{\odot}$.

The modeling technique assumes that the spectra are combinations of multiple bursts of single age, single metallicity stellar populations and fits them with a uniform dust screen model dependent on four parameters: stellar attenuation ($E(B-V)$), the selected reddening curve ($\kappa_{\lambda}$), and linear coefficients ($X_{i}$) of each SB99 model ($M_{i}$) (see equation 1 from C19). The stellar attenuation $E(B-V)$ is allowed to range from 0.0 to 5.0. C19 selects the reddening curve from \cite{Reddy_2016} as it extends closer to the ionizing continua of massive stars ($\sim$ 950\AA) than other models. C19 found that changing the attenuation law to that of \cite{Calzetti_2000} reddens the inferred $E(B-V)$ by 0.01 mag.

The SED shape and observed stellar continuum can be fully described by these four parameters. Though the technique readily allows for more parameter constraints on the SED model, we did not define a free parameter for the absorption caused by the IGM (\teff). To do so, we would have had to subscribe to a predetermined functional form for the redshift evolution of \teff. Instead, we explored the results independent of any such formalism.

Using MPFIT \citep{Markwardt_2009}, an IDL-based, least-squares fitting package\footnote{https://pages.physics.wisc.edu/~craigm/idl/cmpfit.html}, we determined the linear combination of coefficients ($X_{i} \geq 0$) that best describe the observed stellar continuum. The linear coefficients can also be translated to light fractions ($\mathrm{L_{frac}}$) that each model $M_{i}$ contributes to the total intrinsic flux at 1270\AA. Using these light fractions, we can estimate the age and the metallicity of the source (see table \ref{tab:params}). These light-weighted properties of our simple stellar populations are driven by spectral features which are less degenerate than the spectral shape alone. C19 explores the stability of the fitting procedure by measuring the change in flux (per wavelength index) for variations in metallicity and age of model $M_{i}$. Increasing the age of a 0.2 $Z_{\odot}$ model from 2Myr to 8Myr, changed the integrated root square flux of the SED by 2.4 in the 1250-1350\AA\ region. Increasing the metallicity of a 5 Myr model from 0.05$Z_{\odot}$ to 0.4$Z_{\odot}$, changed the integrated root square flux of the SED by 2.1 in the same wavelength region.

The following procedure was used to apply the C19 SED modeling technique to our two LBG composites and 10,000 randomized iterations (we used the 1D error spectra defined in section \ref{subsec:bootstrap} for each redshift bin accordingly):
\begin{itemize}
    \item We masked out 14 ISM absorption lines and non-resonant emissions ($\pm$ 500 \kms) redward of \lya\ that would otherwise contaminate the fitting (see table \ref{tab:ISM})
    
    \item We fit our data in the 1225-1400\AA\ range, to take advantage of the unattenuated sections of our spectra and extrapolated the continuum into the \lya\ forest
    
    \item Using the attenuation curve from \cite{Reddy_2016}, we reddened our fitting results and normalized them in the same range as the composites (1260-1304\AA).
    
    \item We rebinned the SED models to a matching velocity dispersion of 300\kms. See table \ref{tab:params} for the fitted parameters. In the end we were left with the unabsorbed continua of our two composites and those of the 10,000 bootstrap iterations.
\end{itemize}

\begin{table}[ht]
    \centering
    \caption{The transition lines between $\sim 1230-1400\AA$ excluded from our SB99 fitting. Most of these ions are from the ISM \citep{Leitherer_2011} and were masked so that the fit could be extended blueward of \lya. Each line was padded with a $\pm$ 500 \kms buffer.}
    \label{tab:ISM}
    \vskip0.1in
    \begin{tabular}{c|c}
    \hline
    \hline
    Ion & $\lambda_{lab} (\AA)$\\
    \hline
    HI & 1215.67\\
    NV & 1238.82\\
    NV & 1242.80\\
    SiII & 1260.42\\
    SiIII & 1294.54\\
    CIII & 1296.33\\
    SiIII & 1296.74\\
    SiIII & 1298.93\\
    OI & 1302.17\\
    SiII & 1304.37\\
    NiII & 1317.22\\
    CII & 1334.53\\
    CII' & 1335.71\\
    SiIV & 1393.76\\
    SiIV & 1402.77\\
    \hline
    \end{tabular}
\end{table}

\begin{table*}[ht]
\begin{center}
\caption{Best fit parameters and derived values from the SB99 SED modeling for our two composites}
\label{tab:params}
\vskip0.1in
\begin{tabular}{c|c|c|c|c|c}
\hline
\hline
Redshift interval & $N_{spec}$ & $\chi^{2}$ & $E(B-V)$ & Age (Myr) & Metalicity ($Z_\odot$)\\
\hline
$2.43$ (\loz) & 137 & 4.048 & 0.261 & 6.8 & 0.05\\
$2.58$ (\hiz) & 142 & 2.345 & 0.235 & 5.0 & 0.05\\
\hline
\end{tabular}
\end{center}
\end{table*}

\begin{figure*}[ht]
    \begin{center}
    \includegraphics[scale=.5]{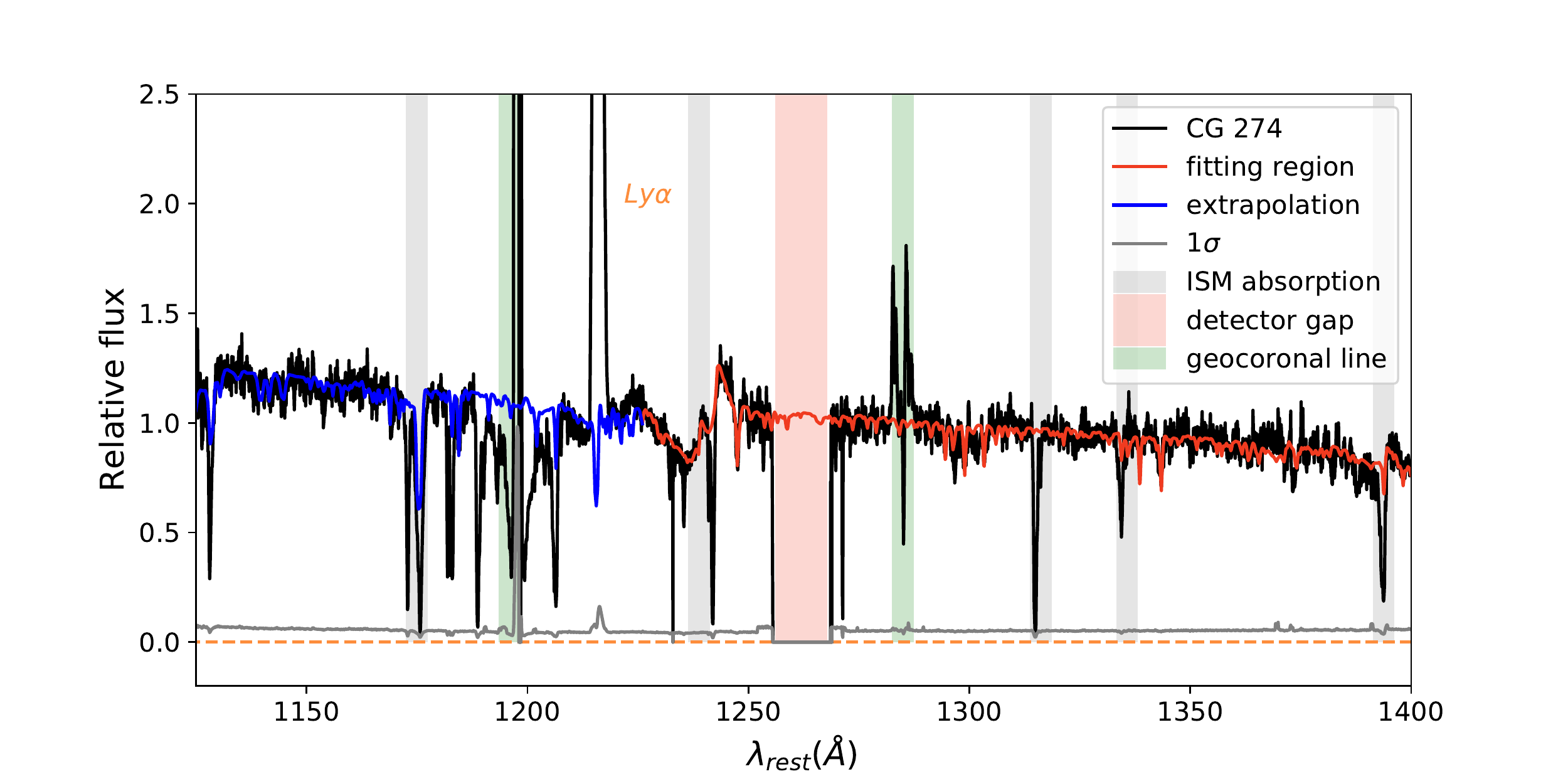}
    \caption{An example of the C19 SB99 fitting routine successfully modeling the continuum shape blueward of \lya. CG 274 was normalized at 1270\AA\ and was fit in the 1223-1406\AA\ region. The SED model extends to $\sim$ 1100\AA. There are two earth-glow sky emission features (shown in green) at 1195\AA\ and 1290\AA\ several ISM absorption lines that are not part of the actual galaxy spectrum. Figure 2 from \citet{Chisholm_2015} also demonstrates similar success at reproducing the blue continuum.}
    \label{fig:sb99}
    \end{center}
\end{figure*}

To demonstrate the stability of our selected fitting technique, we include the fitted SED of a low-$z$ galaxy, CG 274, which has negligible attenuation by the IGM (see figure \ref{fig:sb99}). This spectrum was taken by the Cosmic Origins Spectrograph (COS) on the Hubble Space Telescope using the G130M grating and a central wavelength of 1291\AA\ (Program ID: 15099; PI: Chisholm) . At a redshift $z = 0.0148$, it does not exhibit any notable \lya\ forest absorption. We modelled its flux redward of \lya\ in a similar fashion to the $z \sim 2$ composites and then extrapolated blueward. We find that the extrapolation accurately reproduces the stellar continuum shape, validating our procedure.

\section{\teff\ Measurements \& Associated Errors}
\label{sec:Effective Opacity Measurements}

Armed with an SED model for each composite and bootstrap realization, we analyzed the \lya\ forest to measure an effective opacity at each wavelength index. We used the 1070-1170\AA\ range to avoid continuum fitting problems associated with rapidly changing emission-line profiles, and possible contamination from the proximity effect \citep{Kirkman_2005}. For each stack, we masked the following forest ISM lines with a $\pm$ 5\AA\ buffer: 1083.99, 1117.97, 1122.52, 1128.01, 1144.93, 1152.81\AA. Next, we measured the effective opacity of every stack/model pair for each wavelength index. 

\begin{equation}
        \tau_{\rm eff} = -\ln \frac{F_{\rm obs}}{F_{\rm model}} \;\;
        \label{eq:teff}
\end{equation}
Here, $F_{\rm obs}$ is the average flux and $F_{\rm model}$ is the extrapolated SED. 

\subsection{Metal Corrections}

As we hoped to compare our \teff\ directly to other works that carry out similar analyses \citep{Schaye_2003, Kirkman_2005, Becker_2013}, we corrected our values for absorption from metal lines. Though there are several ways of addressing the contribution to \teff\ from metal absorption (or damped absorbers in the case of quasar spectra) we followed the example of \cite{Kirkman_2005} by subtracting the
metal absorption statistically. We chose to apply this method because their solution did not require identification of contaminating metal lines by eye like that of \cite{Schaye_2003}. Instead \cite{Kirkman_2005} built on a method originally designed by \cite{Tytler_2004} and estimated the metal absorption over the extended redshift range (1.7-3.54) using a sample of 52 quasars.

In general the absorption due to metals in composite spectra essentially scales the mean flux by a relatively minor factor, which becomes increasingly less important at higher redshifts ($z \sim 4$) \citep{Becker_2013}. In fact \cite{Faucher-Giguere_2008} compared the two correction methods from \cite{Schaye_2003} and \cite{Kirkman_2005}, finding that either method was accurate to the level of their statistical error bars.

To find the metal contribution as a function of rest-frame wavelength, we used \cite[equation 1]{Kirkman_2005}. They define DM as the amount of absorption from metal lines alone, as originally coined in \cite{Tytler_2004}. We converted the DM value to $\tau_{M}$ despite the fact that DM is approximately equal to $\tau_{M}$ for z $\sim$ 2.

\begin{equation}
        DM = 0.0156 - (4.646 * 10^{-5})(\lambda_{rest} - 1360\AA)) \;\;
        \label{eq:DM}
\end{equation}

\begin{equation}
        \tau_{M} = ln(1 + DM) \;\;
        \label{eq:tauM}
\end{equation}
Where $\tau_{M}$ is the contribution to the absorption from metals and where $\lambda_{rest}$ is a wavelength index in the forest of the stack. Then, by subtracting the contribution from metals we were left with corrected values of \teff.

\begin{equation}
        \tau_{\rm eff} = \tau_{\rm total} - \tau_{M}\;\;
        \label{eq:corrected}
\end{equation}
Where $\tau_{total}$ is simply the total observed optical depth and \teff\ is the observed optical depth that has been corrected for metal absorption. 

\subsection{Redshift Interval}

Finally, because we were interested in measuring the redshift evolution of \teff\, we converted the wavelength arrays to values of $z$, sampling the entirety of the redshift window included in each stack.

\begin{equation}
        z_{i} = (\lambda_{rest}/1216\AA)(1 + z_{med}) - 1  \;\; 
        \label{eq:zi}
\end{equation}
Where $z_{i}$ is the redshift of a particular absorber in the IGM and \zem\ is the median redshift value of each stack. Because the \zem\ values of our two composite intervals were similar, their redshift coverage overlapped (see appendix table \ref{tab:tau_vals}). Combining both redshift intervals, we were left with a total of 88 indices (56 from the \hiz\ interval and 58 from the \loz\ interval) from which we measured \teff\ in the 1070-1170\AA\ range. This combined redshift sample extended from 2.02 - 2.44, with a median value of 2.22.

\subsection{Error Estimates on \teff}

The errors on the \teff\ values ($\sigma_{\tau}$) were directly measured from the bootstrap analysis but were not simply the standard deviation of each redshift interval across the bootstrap. Instead, we report the diagonals of the covariance matrices in \teff (found using the same method as described in \ref{subsec:bootstrap}). We did not report uncertainty for redshift values as they were only dependent on the \zem\ and $\lambda_{i}$, neither of which had defined errors. For our combined set of measurements and their 1D errors, see appendix table \ref{tab:tau_vals}.

\subsection{Power Law Fitting}

Using a least-squares formalism and the full bootstrap-generated covariance matrices, we fit our combined measurements of \teff\ with the following analytic power-law function.

\begin{equation}
        \tau_{\rm eff} = A[(1+z)/(1+z_{piv})]^{B}  \;\;
        \label{eq:powerlaw}
\end{equation}
where $A$ and $B$ are the scale factor and power-law index parameters. The $z_{piv}$ value included in the fitting function, shifts the power-law index pivot, normalizing the fit to our redshift range \citep{Becker_2013}. We chose $z_{piv} = 2.22$ as it is the median value of our \lya\ forest redshift distribution as measured from \ref{eq:zi}. Our best fit scale factor and power-law index parameters are $A = \Aval \pm \Aerr$ and $B = \Bval \pm \Berr$ respectively. As shown in figure \ref{fig:powerlaw}, the measurements scatter about this curve in a roughly stochastic manner consistent with the uncertainty estimates. One does, however, identify a set of measurements that lie significantly above the model at $z \sim 2.1-2.2$. We attribute these fluctuations to spectral features not smoothed out in our composite spectra. They have not greatly influenced the model because of their small number and significant error estimates.

\subsection{Redshift Evolution in \teff}
\label{subsec:literature}

\begin{figure*}[ht]
    \centering
    \includegraphics[scale=.5]{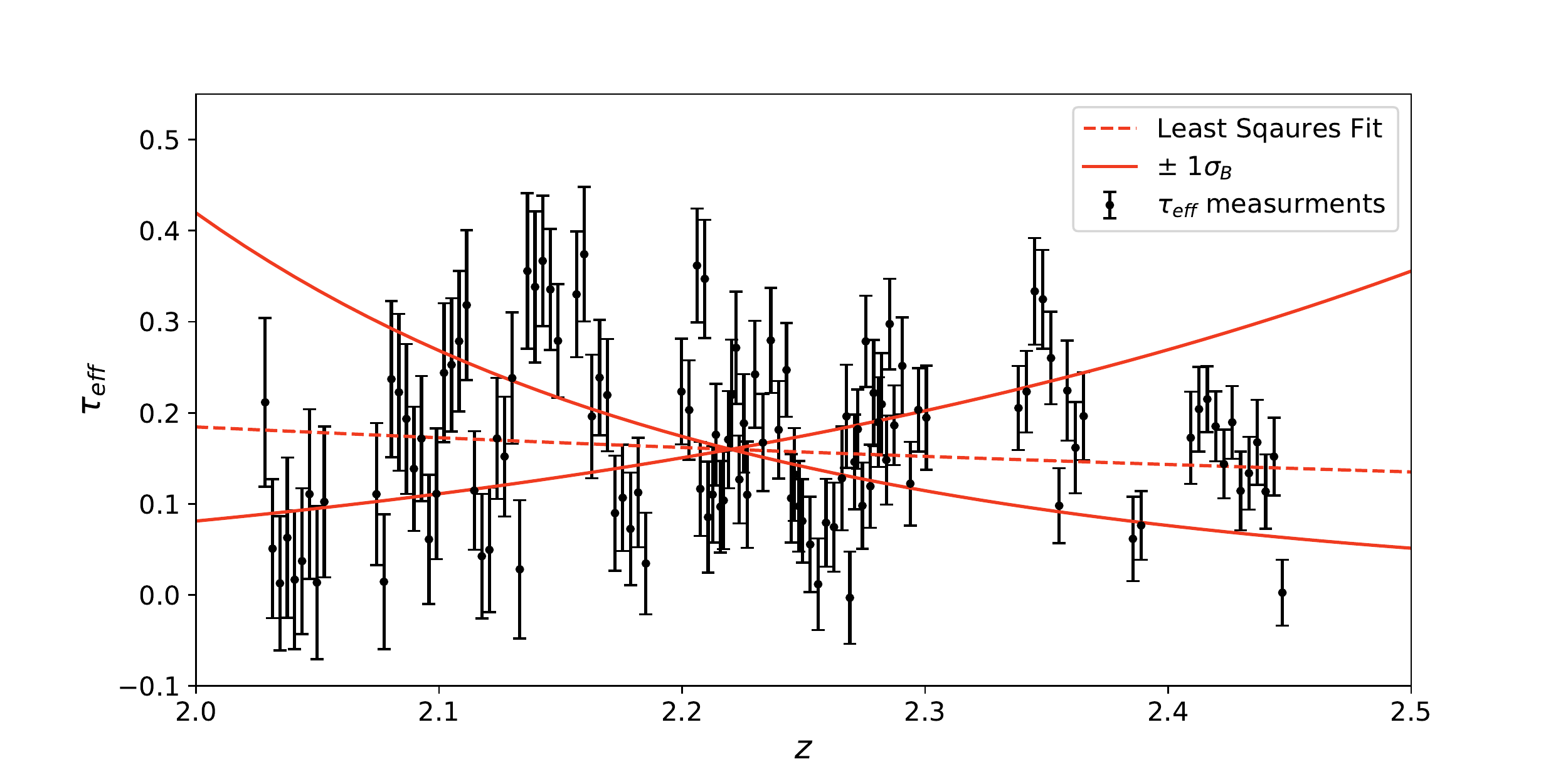}
    \caption{Our measurements of \teff\ as a function of redshift from both redshift intervals. Though we used the full covariance matrix to fit our data with the power law from \ref{eq:powerlaw}, only the diagonals of the matrix are shown. The best fit, is plotted by the dotted line with a 1$\sigma$ uncertainty in the power-law exponent as solid lines.}
    \label{fig:powerlaw}
\end{figure*}

With a best fit power law index error $\sigma_{B}$ = \Berr, we report poor sensitivity to the known evolution of \teff\ at redshifts higher than $z = z_{piv}$. In short, the redshift evolution of \teff\ past $z \sim 3$ was difficult to model given the scatter of our measurements in our narrow redshift window $\Delta z \sim 0.5$ (see figure \ref{fig:powerlaw}).

Evaluating our model at $z = z_{piv}$ we found \teff\ = \Aval $\pm$ \Aerr. This uncertainty does not include a contribution from the error in our power law index parameter. We exluded $\sigma_{B}$ in our error estimate at $z = z_{piv}$ because our data were not sensitive to that parameter. We note that our statistical estimate in the uncertainty of \teff at $z = z_{piv}$ ignores systematic errors which we expect to be at at least 10\%.

Comparing our linear fit's prediction of \teff\ at $z = z_{piv}$ to previous estimates from analysis of quasar spectra; \cite{Kirkman_2005} and \cite{Becker_2013} \teff\ = 0.143, 0.152 (respectively), we found good agreement. We did not find similar compliance with the power law fit from \cite{Schaye_2003} as they predicted \teff = 0.298 at $z = z_{piv}$. This might be because their sample of 21 quasars were significantly contaminated by metal lines, resulting in slight overestimation around $z \sim 2$.

\begin{figure}[ht]
    \centering
    \includegraphics[width=\columnwidth]{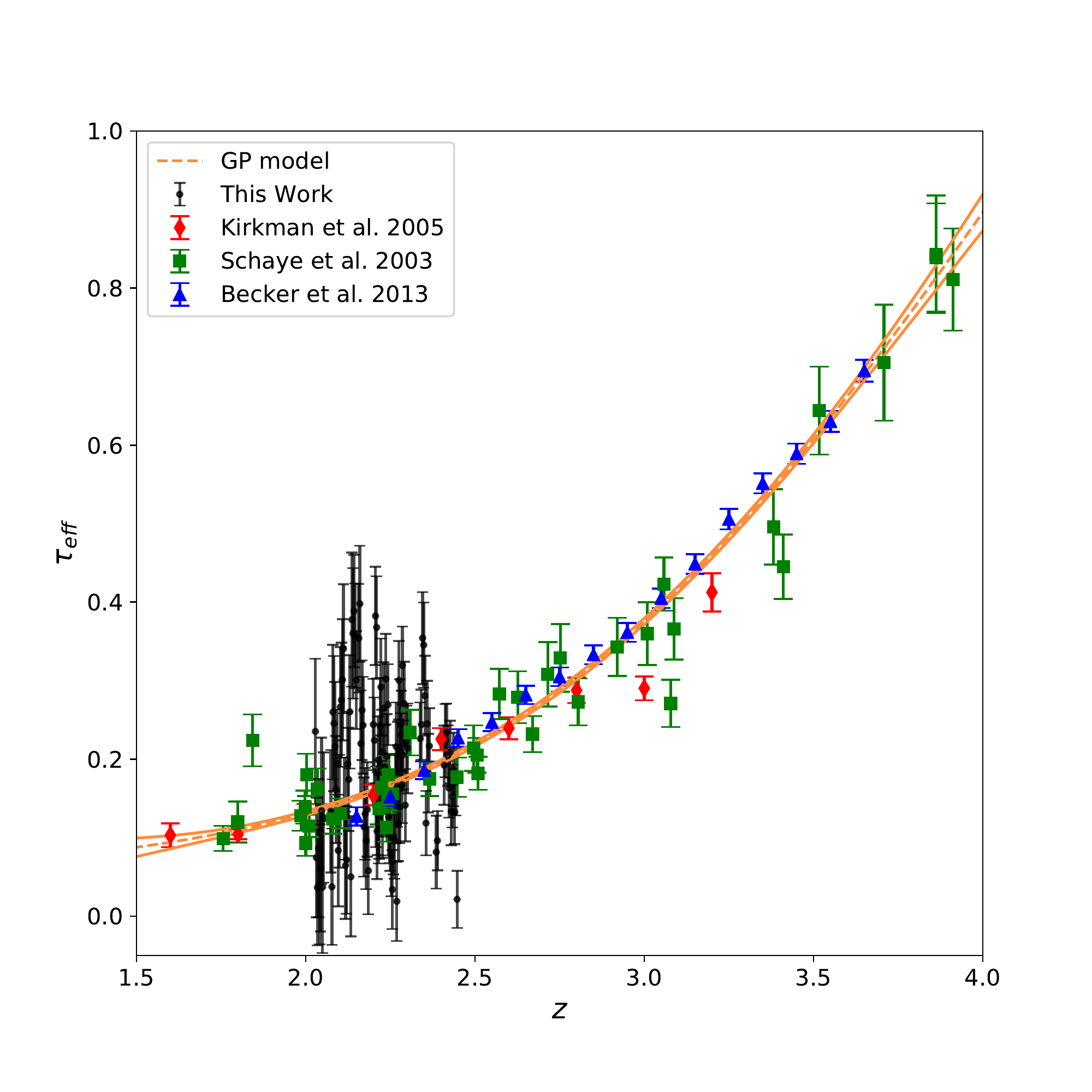}
    \caption{The best fit GP model and associated uncertainty. All of the data points included in the figure \citep{Schaye_2003, Kirkman_2005, Becker_2013} were corrected for intervening metals (and optically thick absorbers in the case of the QSO studies). All analyses shown, were used to constrain the GP model which successfully predicts the steep redshift evolution in \teff.}
    \label{fig:gaussian}
\end{figure}

To further compare results against previous works \citep{Schaye_2003, Kirkman_2005, Becker_2013} and to model the redshift evolution of \teff past $z \sim 3$, we looked to Gaussian Processes (GP). While common practice is to fit such data with a power-law (equation \ref{eq:powerlaw}), recent datasets are not sufficiently well-described by this model \citep{Becker_2013}. Therefore, we analyzed our data alongside the results from \cite{Schaye_2003, Kirkman_2005, Becker_2013} with a GP model which solves for the optimal functional form describing the data. After experimentation, we settled on a Radial Basis Function (RBF) kernel which has mean-square derivatives of all orders and thus creates a smooth fit (see figure \ref{fig:gaussian}). This model is provided by the SciKit Learn toolbox\footnote{https://scikit-learn.org/~gaussian.process.kernels.RBF} \citep{scikit-learn}. We did not fit the GP model with our full covariance matrix nor did we use any of the reported 2D errors for  \citep{Schaye_2003, Kirkman_2005, Becker_2013}. Instead, to simplify the analyses, we only used the 1D diagonals as errors in our measurement.
 
\section{Summary and Concluding Remarks} 
\label{sec:conclusions}

We used 281 LBG spectra collected by the CLAMATO survey to create two composite spectra in the following redshift intervals: $2.25 < z < 2.5$ and $2.5 < z < 2.75$. The normalized composites were fit with simple stellar population SB99 models at rest wavelengths 1225-1400\AA. Extrapolations of these models blueward of \lya\ provided estimates of the effective optical depth \teff\ of the IGM from $z \approx 2.0 -2.5$. We derived bootstrap-generated errors based on the variance in our LBG stacking and propagated these through to the SED fitting.

Our primary results are:

1) A best-fit to the power-law $\mteff = A[(1+z)/(1+2.22)]^{B}$, giving measurements $A = \Aval \pm \Aerr$ and $B = \Bval \pm \Berr$.

2) Our estimate of \teff\ = \Aval $\pm$ \Aerr at $z = 2.22$ is in good agreement with previous estimations based on quasar analysis. This demonstrates that quasar continuum estimations at $z<2.5$ is not subject to large systematic uncertainties.

3) A Gaussian Processes prediction of the redshift evolution of \teff\ using a Radial Basis Kernel. In conjunction with \cite{Schaye_2003, Kirkman_2005, Becker_2013} we show strong evolution in \teff\ at $z > 2$.

As we progress to the next generation of large-scale galaxy surveys at $z>2$ (e.g.\ Prime Focus Spectrograph survey), it is possible that measurements of \teff\ will be drawn primarily from analyses of LBGs. Of course, a continued comparison between quasars and galaxies will
be critical to assess systematic uncertainties associated with continuum estimation.
 
\begin{acknowledgements}

We thank the anonymous referee for careful reading of the original manuscript and their insightful suggestions. JSM would like to thank Marie Lau, Sunil Sihma and Jiani Ding for their input with several aspects of this project. KGL acknowledges support from JSPS KAKENHI Grant Number JP19K14755. The data presented herein were obtained at the W.M. Keck Observatory, which is operated as a scientific partnership among the California Institute of Technology, the University of California and the National Aeronautics and Space Administration (NASA). The Observatory was made possible by the generous financial support of the W.M. Keck Foundation. The authors also wish to recognize and acknowledge the very significant cultural role and reverence that the summit of Maunakea has always had within the indigenous Hawai$'$ian community. We are most fortunate to have the opportunity to conduct observations from this mountain. Support for this work was provided by NASA through the NASA Hubble Fellowship grant \#51432 awarded by the Space Telescope Science Institute, which is operated by the Association of Universities for Research in Astronomy, Inc., for NASA, under contract NAS5-26555. Based on observations made with the NASA/ESA Hubble Space Telescope, obtained from the data archive at the Space Telescope Science Institute. STScI is operated by the Association of Universities for Research in Astronomy, Inc. under NASA contract NAS 5-26555.Support for this work was provided by NASA through grant number 15099 from the Space Telescope Science Institute, which is operated by AURA, Inc., under NASA contract NAS 5-26555.

\end{acknowledgements}

\bibliography{IGM_refs.bib}

\begin{thebibliography}{}
\expandafter\ifx\csname natexlab\endcsname\relax\def\natexlab#1{#1}\fi
\providecommand{\url}[1]{\href{#1}{#1}}
\providecommand{\dodoi}[1]{doi:~\href{http://doi.org/#1}{\nolinkurl{#1}}}
\providecommand{\doeprint}[1]{\href{http://ascl.net/#1}{\nolinkurl{http://ascl.net/#1}}}
\providecommand{\doarXiv}[1]{\href{https://arxiv.org/abs/#1}{\nolinkurl{https://arxiv.org/abs/#1}}}

\bibitem[{Becker {et~al.}(2013)Becker, Hewett, Worseck, \&
  Prochaska}]{Becker_2013}
Becker, G.~D., Hewett, P.~C., Worseck, G., \& Prochaska, J.~X. 2013, Monthly
  Notices of the Royal Astronomical Society, 430, 2067,
  \dodoi{10.1093/mnras/stt031}

\bibitem[{{Becker} {et~al.}(2007){Becker}, {Rauch}, \& {Sargent}}]{Becker_2007}
{Becker}, G.~D., {Rauch}, M., \& {Sargent}, W.~L.~W. 2007, \apj, 662, 72,
  \dodoi{10.1086/517866}

\bibitem[{{Bernardi} {et~al.}(2003){Bernardi}, {Sheth}, {SubbaRao}, {Richards},
  {Burles}, {Connolly}, {Frieman}, {Nichol}, {Schaye}, {Schneider}, {Vanden
  Berk}, {York}, {Brinkmann}, \& {Lamb}}]{Bernardi_2003}
{Bernardi}, M., {Sheth}, R.~K., {SubbaRao}, M., {et~al.} 2003, \aj, 125, 32,
  \dodoi{10.1086/344945}

\bibitem[{Calzetti {et~al.}(2000)Calzetti, Armus, Bohlin, Kinney, Koornneef, \&
  StorchiBergmann}]{Calzetti_2000}
Calzetti, D., Armus, L., Bohlin, R.~C., {et~al.} 2000, The Astrophysical
  Journal, 533, 682, \dodoi{10.1086/308692}

\bibitem[{{Cardelli} {et~al.}(1989){Cardelli}, {Clayton}, \&
  {Mathis}}]{CCM_1989}
{Cardelli}, J.~A., {Clayton}, G.~C., \& {Mathis}, J.~S. 1989, \apj, 345, 245,
  \dodoi{10.1086/167900}

\bibitem[{{Chisholm} {et~al.}(2019){Chisholm}, {Rigby}, {Bayliss}, {Berg},
  {Dahle}, {Gladders}, \& {Sharon}}]{Chisholm_2019}
{Chisholm}, J., {Rigby}, J.~R., {Bayliss}, M., {et~al.} 2019, \apj, 882, 182,
  \dodoi{10.3847/1538-4357/ab3104}

\bibitem[{{Chisholm} {et~al.}(2015){Chisholm}, {Tremonti}, {Leitherer}, {Chen},
  {Wofford}, \& {Lundgren}}]{Chisholm_2015}
{Chisholm}, J., {Tremonti}, C.~A., {Leitherer}, C., {et~al.} 2015, \apj, 811,
  149, \dodoi{10.1088/0004-637X/811/2/149}

\bibitem[{{Dall'Aglio} {et~al.}(2008){Dall'Aglio}, {Wisotzki}, \&
  {Worseck}}]{Dall'Aglio_2008}
{Dall'Aglio}, A., {Wisotzki}, L., \& {Worseck}, G. 2008, \aap, 491, 465,
  \dodoi{10.1051/0004-6361:200810724}

\bibitem[{{de Mello} {et~al.}(2000){de Mello}, {Leitherer}, \&
  {Heckman}}]{de_Mello_2000}
{de Mello}, D.~F., {Leitherer}, C., \& {Heckman}, T.~M. 2000, \apj, 530, 251,
  \dodoi{10.1086/308358}

\bibitem[{Faucher-Giguere {et~al.}(2008)Faucher-Giguere, Prochaska, Lidz,
  Hernquist, \& Zaldarriaga}]{Faucher-Giguere_2008}
Faucher-Giguere, C.-A., Prochaska, J.~X., Lidz, A., Hernquist, L., \&
  Zaldarriaga, M. 2008, The Astrophysical Journal, 681, 831,
  \dodoi{10.1086/588648}

\bibitem[{{Green} {et~al.}(2018){Green}, {Schlafly}, {Finkbeiner}, {Rix},
  {Martin}, {Burgett}, {Draper}, {Flewelling}, {Hodapp}, {Kaiser}, {Kudritzki},
  {Magnier}, {Metcalfe}, {Tonry}, {Wainscoat}, \& {Waters}}]{Green_2018}
{Green}, G.~M., {Schlafly}, E.~F., {Finkbeiner}, D., {et~al.} 2018, \mnras,
  478, 651, \dodoi{10.1093/mnras/sty1008}

\bibitem[{{Gunn} \& {Peterson}(1965)}]{Gunn_Peterson_1965}
{Gunn}, J.~E., \& {Peterson}, B.~A. 1965, \apj, 142, 1633,
  \dodoi{10.1086/148444}

\bibitem[{{Hassan} {et~al.}(2020){Hassan}, {Finlator}, {Dav{\'e}}, {Churchill},
  \& {Prochaska}}]{Hassan_2020}
{Hassan}, S., {Finlator}, K., {Dav{\'e}}, R., {Churchill}, C.~W., \&
  {Prochaska}, J.~X. 2020, \mnras, 492, 2835, \dodoi{10.1093/mnras/staa056}

\bibitem[{Kirkman {et~al.}(2005)Kirkman, Tytler, Suzuki, Melis, Hollywood,
  James, So, Lubin, Jena, Norman, \& Paschos}]{Kirkman_2005}
Kirkman, D., Tytler, D., Suzuki, N., {et~al.} 2005, Monthly Notices of the
  Royal Astronomical Society, 360, 1373,
  \dodoi{10.1111/j.1365-2966.2005.09126.x}

\bibitem[{{Kriek} {et~al.}(2015){Kriek}, {Shapley}, {Reddy}, {Siana}, {Coil},
  {Mobasher}, {Freeman}, {de Groot}, {Price}, {Sanders}, {Shivaei}, {Brammer},
  {Momcheva}, {Skelton}, {van Dokkum}, {Whitaker}, {Aird}, {Azadi}, {Kassis},
  {Bullock}, {Conroy}, {Dav{\'e}}, {Kere{\v{s}}}, \& {Krumholz}}]{Kriek_2015}
{Kriek}, M., {Shapley}, A.~E., {Reddy}, N.~A., {et~al.} 2015, \apjs, 218, 15,
  \dodoi{10.1088/0067-0049/218/2/15}

\bibitem[{{Le Fevre}(2015)}]{LeFevre_2015}
{Le Fevre}, O. 2015, in IAU General Assembly, Vol.~29, 2258239

\bibitem[{Lee {et~al.}(2014)Lee, Hennawi, White, Croft, \& Ozbek}]{Lee_2014}
Lee, K.-G., Hennawi, J.~F., White, M., Croft, R., \& Ozbek, M. 2014, The
  Astrophysical Journal, 788, 49, \dodoi{10.1088/0004-637X/788/1/49}

\bibitem[{Lee {et~al.}(2018)Lee, Krolewski, White, Schlegel, Nugent, Hennawi,
  MÃ¼ller, Pan, Prochaska, Font-Ribera, Suzuki, Glazebrook, Kacprzak,
  Kartaltepe, Koekemoer, FÃ©vre, Lemaux, Maier, Nanayakkara, Rich, Sanders,
  Salvato, Tasca, \& Tran}]{Lee_2018}
Lee, K.-G., Krolewski, A., White, M., {et~al.} 2018, The Astrophysical Journal
  Supplement Series, 237, 31, \dodoi{10.3847/1538-4365/aace58}

\bibitem[{{Leitherer} {et~al.}(2011){Leitherer}, {Tremonti}, {Heckman}, \&
  {Calzetti}}]{Leitherer_2011}
{Leitherer}, C., {Tremonti}, C.~A., {Heckman}, T.~M., \& {Calzetti}, D. 2011,
  \aj, 141, 37, \dodoi{10.1088/0004-6256/141/2/37}

\bibitem[{{Leitherer} {et~al.}(1999){Leitherer}, {Schaerer}, {Goldader},
  {Delgado}, {Robert}, {Kune}, {de Mello}, {Devost}, \&
  {Heckman}}]{Leitherer_1999}
{Leitherer}, C., {Schaerer}, D., {Goldader}, J.~D., {et~al.} 1999, \apjs, 123,
  3, \dodoi{10.1086/313233}

\bibitem[{{Lilly} {et~al.}(2007){Lilly}, {Le F{\`e}vre}, {Renzini}, {Zamorani},
  {Scodeggio}, {Contini}, {Carollo}, {Hasinger}, {Kneib}, {Iovino}, {Le Brun},
  {Maier}, {Mainieri}, {Mignoli}, {Silverman}, {Tasca}, {Bolzonella},
  {Bongiorno}, {Bottini}, {Capak}, {Caputi}, {Cimatti}, {Cucciati}, {Daddi},
  {Feldmann}, {Franzetti}, {Garilli}, {Guzzo}, {Ilbert}, {Kampczyk}, {Kovac},
  {Lamareille}, {Leauthaud}, {Le Borgne}, {McCracken}, {Marinoni}, {Pello},
  {Ricciardelli}, {Scarlata}, {Vergani}, {Sanders}, {Schinnerer}, {Scoville},
  {Taniguchi}, {Arnouts}, {Aussel}, {Bardelli}, {Brusa}, {Cappi}, {Ciliegi},
  {Finoguenov}, {Foucaud}, {Franceschini}, {Halliday}, {Impey}, {Knobel},
  {Koekemoer}, {Kurk}, {Maccagni}, {Maddox}, {Marano}, {Marconi}, {Meneux},
  {Mobasher}, {Moreau}, {Peacock}, {Porciani}, {Pozzetti}, {Scaramella},
  {Schiminovich}, {Shopbell}, {Smail}, {Thompson}, {Tresse}, {Vettolani},
  {Zanichelli}, \& {Zucca}}]{Lilly_2007}
{Lilly}, S.~J., {Le F{\`e}vre}, O., {Renzini}, A., {et~al.} 2007, \apjs, 172,
  70, \dodoi{10.1086/516589}

\bibitem[{{Markwardt}(2009)}]{Markwardt_2009}
{Markwardt}, C.~B. 2009, Astronomical Society of the Pacific Conference Series,
  Vol. 411, {Non-linear Least-squares Fitting in IDL with MPFIT}, ed. D.~A.
  {Bohlender}, D.~{Durand}, \& P.~{Dowler}, 251

\bibitem[{McDonald {et~al.}(2006)McDonald, Seljak, Burles, Schlegel, Weinberg,
  Shih, Schaye, Schneider, Brinkmann, Brunner, \& Fukugita}]{Mcdonald_2006}
McDonald, P., Seljak, U., Burles, S., {et~al.} 2006, The Astrophysical Journal
  Supplement Series, 163, 80, \dodoi{10.1086/444361}

\bibitem[{McQuinn(2016)}]{McQuinn_2016}
McQuinn, M. 2016, Annual Review of Astronomy and Astrophysics, 54, 313–362,
  \dodoi{10.1146/annurev-astro-082214-122355}

\bibitem[{Meiksin(2009)}]{Meiksin_2009}
Meiksin, A.~A. 2009, Reviews of Modern Physics, 81, 1405,
  \dodoi{10.1103/RevModPhys.81.1405}

\bibitem[{{Nanayakkara} {et~al.}(2016){Nanayakkara}, {Glazebrook}, {Kacprzak},
  {Yuan}, {Tran}, {Spitler}, {Kewley}, {Straatman}, {Cowley}, {Fisher},
  {Labbe}, {Tomczak}, {Allen}, \& {Alcorn}}]{Nanayakkara_2016}
{Nanayakkara}, T., {Glazebrook}, K., {Kacprzak}, G.~G., {et~al.} 2016, \apj,
  828, 21, \dodoi{10.3847/0004-637X/828/1/21}

\bibitem[{{O'Donnell}(1994)}]{O'Donnell_1994}
{O'Donnell}, J.~E. 1994, \apj, 422, 158, \dodoi{10.1086/173713}

\bibitem[{{Oke} {et~al.}(1995){Oke}, {Cohen}, {Carr}, {Cromer}, {Dingizian},
  {Harris}, {Labrecque}, {Lucinio}, {Schaal}, {Epps}, \& {Miller}}]{LRIS}
{Oke}, J.~B., {Cohen}, J.~G., {Carr}, M., {et~al.} 1995, \pasp, 107, 375,
  \dodoi{10.1086/133562}

\bibitem[{Paris {et~al.}(2011)Paris, Petitjean, Rollinde, Aubourg, Busca,
  Charlassier, Delubac, Hamilton, Le~Goff, Palanque-Delabrouille, Peirani,
  Pichon, Rich, Vargas-MagaÃ±a, \& Yche}]{Paris_2011}
Paris, I., Petitjean, P., Rollinde, E., {et~al.} 2011, Astronomy \&
  Astrophysics, 530, A50, \dodoi{10.1051/0004-6361/201016233}

\bibitem[{Pedregosa {et~al.}(2011)Pedregosa, Varoquaux, Gramfort, Michel,
  Thirion, Grisel, Blondel, Prettenhofer, Weiss, Dubourg, Vanderplas, Passos,
  Cournapeau, Brucher, Perrot, \& Duchesnay}]{scikit-learn}
Pedregosa, F., Varoquaux, G., Gramfort, A., {et~al.} 2011, Journal of Machine
  Learning Research, 12, 2825

\bibitem[{Prochaska {et~al.}(2009)Prochaska, Worseck, \&
  O'Meara}]{Prochaska_2009}
Prochaska, J.~X., Worseck, G., \& O'Meara, J.~M. 2009, The Astrophysical
  Journal, 705, L113, \dodoi{10.1088/0004-637X/705/2/L113}

\bibitem[{Rauch(1998)}]{Rauch_1998}
Rauch, M. 1998, Annual Review of Astronomy and Astrophysics, 36, 267,
  \dodoi{10.1146/annurev.astro.36.1.267}

\bibitem[{Reddy {et~al.}(2008)Reddy, Steidel, Pettini, Adelberger, Shapley,
  Erb, \& Dickinson}]{Reddy_2008}
Reddy, N.~A., Steidel, C.~C., Pettini, M., {et~al.} 2008, The Astrophysical
  Journal Supplement Series, 175, 48, \dodoi{10.1086/521105}

\bibitem[{Reddy {et~al.}(2016)Reddy, Steidel, Pettini, \&
  Bogosavljevic}]{Reddy_2016}
Reddy, N.~A., Steidel, C.~C., Pettini, M., \& Bogosavljevic, M. 2016, The
  Astrophysical Journal, 828, 107, \dodoi{10.3847/0004-637X/828/2/107}

\bibitem[{{Rix} {et~al.}(2004){Rix}, {Pettini}, {Leitherer}, {Bresolin},
  {Kudritzki}, \& {Steidel}}]{Rix_2004}
{Rix}, S.~A., {Pettini}, M., {Leitherer}, C., {et~al.} 2004, \apj, 615, 98,
  \dodoi{10.1086/424031}

\bibitem[{Schaye {et~al.}(2003)Schaye, Aguirre, Kim, Theuns, Rauch, \&
  Sargent}]{Schaye_2003}
Schaye, J., Aguirre, A., Kim, T., {et~al.} 2003, The Astrophysical Journal,
  596, 768, \dodoi{10.1086/378044}

\bibitem[{Scoville {et~al.}(2007)Scoville, Aussel, Brusa, Capak, Carollo,
  Elvis, Giavalisco, Guzzo, Hasinger, Impey, Kneib, LeFevre, Lilly, Mobasher,
  Renzini, Rich, Sanders, Schinnerer, Schminovich, Shopbell, Taniguchi, \&
  Tyson}]{Scoville_2007}
Scoville, N., Aussel, H., Brusa, M., {et~al.} 2007, The Astrophysical Journal
  Supplement Series, 172, 1, \dodoi{10.1086/516585}

\bibitem[{Shapley {et~al.}(2003)Shapley, Steidel, Pettini, \&
  Adelberger}]{Shapley_2003}
Shapley, A.~E., Steidel, C.~C., Pettini, M., \& Adelberger, K.~L. 2003, The
  Astrophysical Journal, 588, 65, \dodoi{10.1086/373922}

\bibitem[{Steidel {et~al.}(1996)Steidel, Giavalisco, Pettini, Dickinson, \&
  Adelberger}]{Steidel_1996}
Steidel, C.~C., Giavalisco, M., Pettini, M., Dickinson, M., \& Adelberger,
  K.~L. 1996, 462, 7

\bibitem[{{Theuns} {et~al.}(2002){Theuns}, {Bernardi}, {Frieman}, {Hewett},
  {Schaye}, {Sheth}, \& {Subbarao}}]{Theuns_2002}
{Theuns}, T., {Bernardi}, M., {Frieman}, J., {et~al.} 2002, \apjl, 574, L111,
  \dodoi{10.1086/342531}

\bibitem[{Thomas {et~al.}(2017)Thomas, Le~Fevre, Le~Brun, Cassata, Garilli,
  Lemaux, Maccagni, Pentericci, Tasca, Zamorani, Zucca, Amorin, Bardelli,
  CassarÃ , Castellano, Cimatti, Cucciati, Durkalec, Fontana, Giavalisco,
  Grazian, Hathi, Ilbert, Paltani, Pforr, Ribeiro, Schaerer, Scodeggio,
  Sommariva, Talia, Tresse, Vanzella, Vergani, Capak, Charlot, Contini, Cuby,
  de~la Torre, Dunlop, Fotopoulou, Koekemoer, LÃ³pez-Sanjuan, Mellier,
  Salvato, Scoville, Taniguchi, \& Wang}]{Thomas_2017}
Thomas, R., Le~Fevre, O., Le~Brun, V., {et~al.} 2017, Astronomy \&
  Astrophysics, 597, A88, \dodoi{10.1051/0004-6361/201425342}

\bibitem[{{Tytler} {et~al.}(2004){Tytler}, {O'Meara}, {Suzuki}, {Kirkman},
  {Lubin}, \& {Orin}}]{Tytler_2004}
{Tytler}, D., {O'Meara}, J.~M., {Suzuki}, N., {et~al.} 2004, \aj, 128, 1058,
  \dodoi{10.1086/423293}

\bibitem[{Worseck {et~al.}(2014)Worseck, Prochaska, O'Meara, Becker, Ellison,
  Lopez, Meiksin, MÃ©nard, Murphy, \& Fumagalli}]{Worseck_2014}
Worseck, G., Prochaska, J.~X., O'Meara, J.~M., {et~al.} 2014, Monthly Notices
  of the Royal Astronomical Society, 445, 1745, \dodoi{10.1093/mnras/stu1827}

\end{thebibliography}

\appendix

\begin{center}
\begin{longtable}{l|c|c|c}
\caption{The sample of CLAMATO LBGs used in our composite analysis} \label{tab:clamato} \\
\hline \multicolumn{1}{|l|}{\textbf{CLAMATO ID}} & \multicolumn{1}{c|}{\textbf{RA}} & \multicolumn{1}{c|}{\textbf{DEC}} & \multicolumn{1}{c|}{\textbf{z}} \\ \hline 
\endfirsthead
\multicolumn{4}{c}
{{\bfseries \tablename\ \thetable{}}} \\
\hline \multicolumn{1}{|l|}{\textbf{CLAMATO ID}} & \multicolumn{1}{c|}{\textbf{RA}} & \multicolumn{1}{c|}{\textbf{DEC}} & \multicolumn{1}{c|}{\textbf{z}} \\ \hline 
\endhead
\hline \multicolumn{4}{|r|}{{Continued on next page}} \\ \hline
\endfoot
\hline \hline
\endlastfoot
cl2016comb-zsp2.3-00871 & 150.08844 & 2.24847 & 2.301 \\
cl2016comb-zsp2.6-00923 & 150.0679 & 2.15819 & 2.621 \\
cl2016comb-zsp2.5-00941 & 150.03569 & 2.2896 & 2.45 \\
cl2016comb-zsp2.7-00954 & 150.02919 & 2.25323 & 2.66 \\
cl2016comb-zsp2.4-01012 & 150.05318 & 2.1513 & 2.516 \\
cl2016comb-zsp2.6-01016 & 150.02277 & 2.14595 & 2.624 \\
cl2016comb-zsp2.4-01321 & 150.02322 & 2.37721 & 2.384 \\
cl2016comb-zsp2.7-01349 & 150.0231 & 2.31791 & 2.675 \\
cl2016comb-zsp2.6-01865 & 150.1011 & 2.24173 & 2.647 \\
cl2016comb-zph2.5-12541 & 150.10332 & 2.2585 & 2.438 \\
cl2016comb-zph2.6-12722 & 150.09888 & 2.16134 & 2.416 \\
cl2016comb-zph2.3-12836 & 150.04671 & 2.25102 & 2.284 \\
cl2016comb-zph2.6-15035 & 150.09714 & 2.45167 & 2.479 \\
cl2016comb-zph2.4-15059 & 150.16531 & 2.42249 & 2.506 \\
cl2016comb-zph2.3-15171 & 150.09419 & 2.34853 & 2.273 \\
cl2016comb-zph2.6-15218 & 150.17313 & 2.3254 & 2.613 \\
cl2016comb-zsp2.6-15363 & 149.98645 & 2.37884 & 2.545 \\
cl2016comb-zsp2.4-15373 & 150.00044 & 2.37243 & 2.42 \\
cl2016comb-zph2.4-15473 & 150.04306 & 2.31694 & 2.44 \\
pc06-zph2.3-15159 & 150.09453 & 2.35827 & 2.466 \\
pc06-zsp2.4-00852 & 150.06163 & 2.28314 & 2.377 \\
cpilot06-zsp2.7-00857 & 150.09343 & 2.27371 & 2.65 \\
cpilot06-zsp2.7-01260 & 150.07938 & 2.3406 & 2.679 \\
cpilot06-zsp2.7-01276 & 150.0798 & 2.30685 & 2.679 \\
cpilot06-zsp2.7-01324 & 150.03629 & 2.37356 & 2.73 \\
cpilot05-zph2.3-12714 & 150.0827 & 2.16487 & 2.26 \\
cpilot02-zph2.5-12826 & 150.00772 & 2.24664 & 2.525 \\
cpilot02-zph2.5-12988 & 149.98288 & 2.1657 & 2.42 \\
cpilot02-zsp2.3-00962 & 150.00296 & 2.24145 & 2.267 \\
cpilot02-zsp2.3-01013 & 149.96033 & 2.15784 & 2.297 \\
cpilot02-zsp2.4-00965 & 149.99504 & 2.2398 & 2.442 \\
cpilot02-zsp2.4-01882 & 149.99516 & 2.23734 & 2.45 \\
cpilot02-zsp2.5-00990 & 149.98834 & 2.20705 & 2.458 \\
cpilot02-zsp2.6-00986 & 149.99481 & 2.21234 & 2.556 \\
cpilot02-zsp2.6-01009 & 150.0136 & 2.16877 & 2.623 \\
cpilot02-zsp2.7-00982 & 150.02107 & 2.21256 & 2.658 \\
cpilot09-zph2.5-15182 & 150.12419 & 2.34884 & 2.513 \\
cpilot09-zph2.5-15268 & 150.15688 & 2.30079 & 2.505 \\
cpilot09-zph2.6-12505 & 150.21675 & 2.36974 & 2.408 \\
cpilot09-zph2.6-15214 & 150.11501 & 2.3276 & 2.551 \\
cpilot09-zph2.7-15220 & 150.12335 & 2.32413 & 2.623 \\
cpilot09-zsp2.5-00856 & 150.161 & 2.2759 & 2.504 \\
cpilot09-zsp2.5-01753 & 150.15979 & 2.37123 & 2.458 \\
cpilot09-zsp2.5-01754 & 150.14763 & 2.36719 & 2.452 \\
cpilot09-zsp2.6-01252 & 150.16002 & 2.35477 & 2.556 \\
cpilot09-zsp2.6-01262 & 150.11871 & 2.33762 & 2.552 \\
cpilot09-zsp2.7-00858 & 150.14117 & 2.27234 & 2.747 \\
cpilot08-zph2.2-12568 & 150.16913 & 2.23838 & 2.451 \\
cpilot08-zph2.3-01886 & 150.21675 & 2.36974 & 2.305 \\
cpilot08-zph2.5-12604 & 150.1651 & 2.22747 & 2.437 \\
cpilot08-zsp2.3-00892 & 150.10474 & 2.21573 & 2.324 \\
cpilot08-zsp2.4-00877 & 150.12111 & 2.23542 & 2.432 \\
cpilot08-zsp2.7-00889 & 150.14442 & 2.21977 & 2.702 \\
cpilot08-zsp2.7-00903 & 150.1205 & 2.1923 & 2.688 \\
cpilot08-zsp2.7-00933 & 150.10455 & 2.13738 & 2.69 \\
cpilot03-zph2.4-15492 & 149.95932 & 2.30758 & 2.555 \\
cpilot03-zph2.6-12812 & 150.0199 & 2.26976 & 2.42 \\
cpilot03-zsp2.3-01330 & 149.99364 & 2.36083 & 2.256 \\
cpilot03-zsp2.5-01345 & 150.0118 & 2.32297 & 2.467 \\
cpilot03-zsp2.6-01352 & 150.01968 & 2.31087 & 2.624 \\
cpilot12-zph2.4-14888 & 150.24263 & 2.35848 & 2.278 \\
cpilot12-zph2.4-14925 & 150.23189 & 2.33713 & 2.456 \\
cpilot12-zph2.4-15146 & 150.22118 & 2.37094 & 2.52 \\
cpilot12-zph2.6-12247 & 150.24257 & 2.27782 & 2.525 \\
cpilot12-zph2.6-14947 & 150.2346 & 2.33237 & 2.505 \\
cpilot12-zph2.6-15161 & 150.22186 & 2.36248 & 2.5 \\
cpilot12-zph2.6-15173 & 150.22505 & 2.35619 & 2.507 \\
cpilot12-zsp2.5-01268 & 150.21138 & 2.32292 & 2.46 \\
cpilot12-zsp2.5-01274 & 150.22343 & 2.3072 & 2.491 \\
cpilot12-zsp2.7-01272 & 150.19978 & 2.3155 & 2.738 \\
npc05-zph2.3-12595 & 150.07341 & 2.23328 & 2.303 \\
npc05-zph2.3-12701 & 150.07675 & 2.17348 & 2.486 \\
npc05-zph2.5-12653 & 150.06866 & 2.18897 & 2.3 \\
npc05-zsp2.3-00964 & 150.05905 & 2.24059 & 2.283 \\
npc05-zsp2.4-01861 & 150.08061 & 2.24284 & 2.437 \\
c16-24-zph2.4-15103 & 150.22504 & 2.39841 & 2.645 \\
c16-24-zph2.4-15121 & 150.2166 & 2.37826 & 2.373 \\
c16-24-zph2.6-17723 & 150.21797 & 2.49178 & 2.645 \\
c16-24-zph2.7-17497 & 150.22946 & 2.47711 & 2.66 \\
c16-24-zsp2.5-01778 & 150.19771 & 2.48577 & 2.49 \\
c16-24-zsp2.6-01219 & 150.20456 & 2.45545 & 2.58 \\
c16-24-zsp2.7-01779 & 150.20668 & 2.48269 & 2.676 \\
c16-11-zph2.4-12707 & 150.20447 & 2.17102 & 2.37 \\
c16-11-zph2.5-12304 & 150.22772 & 2.23602 & 2.493 \\
c16-11-zph2.5-12634 & 150.18759 & 2.20976 & 2.48 \\
c16-11-zsp2.3-00873 & 150.19859 & 2.24642 & 2.367 \\
c16-11-zsp2.4-00884 & 150.20885 & 2.22566 & 2.437 \\
c16-11-zsp2.5-00834 & 150.23257 & 2.14658 & 2.638 \\
c16-20-zph2.6-17715 & 150.09602 & 2.49511 & 2.536 \\
c16-20-zph2.6-26486 & 150.06027 & 2.3877 & 2.55 \\
c16-20-zsp2.5-01241 & 150.07576 & 2.38064 & 2.466 \\
c16-20-zsp2.7-01233 & 150.08762 & 2.39438 & 2.702 \\
c16-22-zph2.4-15040 & 150.13757 & 2.44066 & 2.51 \\
c16-22-zph2.6-17758 & 150.10231 & 2.47219 & 2.606 \\
c16-22-zsp2.5-01239 & 150.14885 & 2.38391 & 2.505 \\
c16-18-zph2.6-15288 & 149.95987 & 2.45353 & 2.515 \\
c16-18-zph2.6-15292 & 149.95877 & 2.45022 & 2.627 \\
c16-18-zsp2.4-01589 & 150.01366 & 2.46674 & 2.417 \\
cl2017comb-zsp2.5-00834 & 150.23257 & 2.14658 & 2.637 \\
cl2017comb-zsp2.3-00871 & 150.08844 & 2.24847 & 2.301 \\
cl2017comb-zsp2.6-00923 & 150.0679 & 2.15819 & 2.622 \\
cl2017comb-zsp2.7-00954 & 150.02919 & 2.25323 & 2.657 \\
cl2017comb-zsp2.6-00966 & 150.03355 & 2.23549 & 2.555 \\
cl2017comb-zsp2.4-01003 & 150.05382 & 2.185 & 2.56 \\
cl2017comb-zsp2.4-01012 & 150.05318 & 2.1513 & 2.452 \\
cl2017comb-zsp2.6-01016 & 150.02277 & 2.14595 & 2.623 \\
cl2017comb-zsp2.3-01181 & 150.33495 & 2.36654 & 2.315 \\
cl2017comb-zsp2.5-01239 & 150.14885 & 2.38391 & 2.507 \\
cl2017comb-zsp2.5-01245 & 150.1015 & 2.37672 & 2.464 \\
cl2017comb-zsp2.4-01265 & 150.06456 & 2.32904 & 2.447 \\
cl2017comb-zsp2.4-01321 & 150.02322 & 2.37721 & 2.376 \\
cl2017comb-zsp2.7-01349 & 150.0231 & 2.31791 & 2.678 \\
cl2017comb-zsp2.6-01865 & 150.1011 & 2.24173 & 2.646 \\
cl2017comb-zph2.5-12541 & 150.10333 & 2.25851 & 2.437 \\
cl2017comb-zph2.6-12722 & 150.09888 & 2.16134 & 2.417 \\
cl2017comb-zsp2.3-12836 & 150.04671 & 2.25102 & 2.284 \\
cl2017comb-zph2.5-14852 & 150.3867 & 2.37505 & 2.456 \\
cl2017comb-zph2.6-15035 & 150.09714 & 2.45167 & 2.479 \\
cl2017comb-zsp2.5-15059 & 150.16531 & 2.42249 & 2.506 \\
cl2017comb-zph2.3-15171 & 150.09421 & 2.34853 & 2.274 \\
cl2017comb-zph2.6-15218 & 150.17313 & 2.3254 & 2.613 \\
cl2017comb-zsp2.4-15373 & 150.00044 & 2.37243 & 2.418 \\
cl2017comb-zph2.7-15399 & 150.0201 & 2.35363 & 2.689 \\
cl2017comb-zph2.4-15473 & 150.04306 & 2.31694 & 2.44 \\
cl2017comb-zsp2.6-15492 & 149.95932 & 2.30758 & 2.555 \\
cl2017comb-zsp2.6-17758 & 150.10231 & 2.47219 & 2.606 \\
pc06-zph2.3-15159 & 150.09453 & 2.35827 & 2.461 \\
pc06-zsp2.4-00852 & 150.06163 & 2.28314 & 2.375 \\
cpilot06-zsp2.7-00857 & 150.09343 & 2.27371 & 2.65 \\
cpilot06-zsp2.7-01260 & 150.07938 & 2.3406 & 2.679 \\
cpilot06-zsp2.7-01276 & 150.0798 & 2.30685 & 2.679 \\
cpilot05-zph2.3-12714 & 150.0827 & 2.16487 & 2.26 \\
cpilot05-zsp2.7-00994 & 150.04597 & 2.20114 & 2.709 \\
cpilot02-zph2.5-12826 & 150.00772 & 2.24664 & 2.525 \\
cpilot02-zph2.5-12988 & 149.98288 & 2.1657 & 2.42 \\
cpilot02-zsp2.3-00962 & 150.00296 & 2.24145 & 2.267 \\
cpilot02-zsp2.3-01013 & 149.96033 & 2.15784 & 2.297 \\
cpilot02-zsp2.4-00965 & 149.99504 & 2.2398 & 2.442 \\
cpilot02-zsp2.4-01882 & 149.99516 & 2.23734 & 2.45 \\
cpilot02-zsp2.5-00990 & 149.98834 & 2.20705 & 2.458 \\
cpilot02-zsp2.6-00986 & 149.99481 & 2.21234 & 2.556 \\
cpilot02-zsp2.6-01009 & 150.0136 & 2.16877 & 2.623 \\
cpilot02-zsp2.7-00982 & 150.02107 & 2.21256 & 2.658 \\
cpilot09-zph2.5-15182 & 150.12419 & 2.34884 & 2.513 \\
cpilot09-zph2.5-15268 & 150.15688 & 2.30079 & 2.502 \\
cpilot09-zph2.6-12505 & 150.14354 & 2.28177 & 2.408 \\
cpilot09-zph2.6-15214 & 150.11501 & 2.3276 & 2.552 \\
cpilot09-zph2.7-15220 & 150.12335 & 2.32413 & 2.623 \\
cpilot09-zsp2.5-00856 & 150.161 & 2.2759 & 2.504 \\
cpilot09-zsp2.5-01753 & 150.15979 & 2.37123 & 2.46 \\
cpilot09-zsp2.5-01754 & 150.14763 & 2.36719 & 2.455 \\
cpilot09-zsp2.6-01252 & 150.16002 & 2.35477 & 2.556 \\
cpilot09-zsp2.6-01262 & 150.11871 & 2.33762 & 2.552 \\
cpilot09-zsp2.7-00858 & 150.14117 & 2.27234 & 2.747 \\
cpilot08-zph2.2-12568 & 150.16913 & 2.23838 & 2.451 \\
cpilot08-zph2.3-01886 & 150.12947 & 2.2072 & 2.305 \\
cpilot08-zph2.5-12604 & 150.1651 & 2.22747 & 2.437 \\
cpilot08-zsp2.3-00892 & 150.10474 & 2.21573 & 2.321 \\
cpilot08-zsp2.4-00877 & 150.12111 & 2.23543 & 2.432 \\
cpilot08-zsp2.7-00889 & 150.14442 & 2.21977 & 2.71 \\
cpilot08-zsp2.7-00903 & 150.1205 & 2.1923 & 2.685 \\
cpilot08-zsp2.7-00933 & 150.10455 & 2.13738 & 2.69 \\
cpilot03-zsp2.3-01330 & 149.99364 & 2.36083 & 2.256 \\
cpilot03-zsp2.7-00951 & 150.0182 & 2.25944 & 2.673 \\
cpilot12-zph2.4-14888 & 150.24263 & 2.35848 & 2.278 \\
cpilot12-zph2.4-14925 & 150.23189 & 2.33713 & 2.456 \\
cpilot12-zph2.4-15146 & 150.22118 & 2.37094 & 2.517 \\
cpilot12-zph2.6-12247 & 150.24257 & 2.27782 & 2.525 \\
cpilot12-zph2.6-14947 & 150.2346 & 2.33237 & 2.505 \\
cpilot12-zph2.6-15161 & 150.22186 & 2.36248 & 2.5 \\
cpilot12-zph2.6-15173 & 150.22505 & 2.35619 & 2.507 \\
cpilot12-zsp2.5-01268 & 150.21138 & 2.32292 & 2.46 \\
cpilot12-zsp2.5-01274 & 150.22343 & 2.3072 & 2.491 \\
cpilot12-zsp2.5-01678 & 150.22528 & 2.3512 & 2.484 \\
cpilot12-zsp2.7-01272 & 150.19978 & 2.3155 & 2.738 \\
npc05-zph2.3-12595 & 150.07341 & 2.23328 & 2.303 \\
npc05-zph2.3-12701 & 150.07675 & 2.17348 & 2.486 \\
npc05-zph2.5-12653 & 150.06866 & 2.18897 & 2.3 \\
npc05-zsp2.3-00964 & 150.05905 & 2.24059 & 2.281 \\
npc05-zsp2.4-01861 & 150.08061 & 2.24284 & 2.437 \\
c16-24-zph2.4-15103 & 150.22504 & 2.39841 & 2.642 \\
c16-24-zph2.4-15121 & 150.2166 & 2.37826 & 2.375 \\
c16-24-zph2.6-17723 & 150.21797 & 2.49178 & 2.645 \\
c16-24-zph2.7-17497 & 150.22946 & 2.47711 & 2.628 \\
c16-24-zsp2.5-01778 & 150.19771 & 2.48577 & 2.535 \\
c16-24-zsp2.6-01219 & 150.20456 & 2.45545 & 2.586 \\
c16-24-zsp2.7-01779 & 150.20668 & 2.48269 & 2.675 \\
c16-11-zph2.3-12434 & 150.23575 & 2.1661 & 2.31 \\
c16-11-zph2.3-12690 & 150.21985 & 2.17724 & 2.61 \\
c16-11-zph2.4-12707 & 150.20447 & 2.17102 & 2.37 \\
c16-11-zph2.5-12304 & 150.22772 & 2.23602 & 2.489 \\
c16-11-zph2.5-12634 & 150.18759 & 2.20976 & 2.48 \\
c16-11-zsp2.4-00884 & 150.20885 & 2.22566 & 2.438 \\
c16-20-zph2.6-17715 & 150.09602 & 2.49511 & 2.538 \\
c16-20-zph2.6-26486 & 150.06027 & 2.3877 & 2.55 \\
c16-20-zsp2.5-01241 & 150.07576 & 2.38064 & 2.469 \\
c16-20-zsp2.7-01233 & 150.08762 & 2.39438 & 2.703 \\
c16-22-zph2.4-15040 & 150.13757 & 2.44066 & 2.51 \\
c16-18-zph2.6-15288 & 149.95987 & 2.45353 & 2.515 \\
c16-18-zph2.6-15292 & 149.95877 & 2.45022 & 2.627 \\
c16-18-zsp2.4-01589 & 150.01366 & 2.46674 & 2.415 \\
c16-18-zsp2.5-01298 & 149.97008 & 2.43493 & 2.458 \\
c17-27s-zph2.4-12455 & 150.24768 & 2.15066 & 2.294 \\
c17-27s-zph2.5-12355 & 150.25565 & 2.21225 & 2.578 \\
c17-27s-zph2.5-32293 & 150.2847 & 2.21318 & 2.503 \\
c17-27s-zph2.6-12374 & 150.27167 & 2.20637 & 2.615 \\
c17-27s-zph2.7-12405 & 150.27063 & 2.18604 & 2.58 \\
c17-27s-zph2.7-32286 & 150.27771 & 2.21997 & 2.495 \\
c17-27s-zsp2.3-00805 & 150.30594 & 2.19577 & 2.323 \\
c17-27s-zsp2.5-00785 & 150.27141 & 2.24478 & 2.506 \\
c17-27s-zsp2.6-00783 & 150.28088 & 2.24953 & 2.579 \\
c17-27s-zsp2.6-00793 & 150.27214 & 2.2301 & 2.611 \\
c17-27s-zsp2.6-00823 & 150.26257 & 2.16603 & 2.601 \\
c17-28s-zph2.6-12252 & 150.30026 & 2.27421 & 2.581 \\
c17-28s-zph2.6-14978 & 150.29825 & 2.31653 & 2.576 \\
c17-28s-zsp2.4-01216 & 150.26845 & 2.2975 & 2.408 \\
c17-28s-zsp2.5-00771 & 150.27863 & 2.27316 & 2.53 \\
c17-28s-zsp2.5-01189 & 150.29594 & 2.3454 & 2.465 \\
c17-28s-zsp2.5-01193 & 150.30426 & 2.33754 & 2.448 \\
c17-28s-zsp2.5-01201 & 150.25424 & 2.33063 & 2.468 \\
c17-28s-zsp2.5-01203 & 150.25378 & 2.32426 & 2.468 \\
c17-29-zph2.3-33410 & 150.28709 & 2.41177 & 2.312 \\
c17-29-zph2.5-33398 & 150.29295 & 2.42088 & 2.402 \\
c17-29-zph2.5-33402 & 150.30779 & 2.41704 & 2.545 \\
c17-29-zph2.6-14815 & 150.30571 & 2.39347 & 2.47 \\
c17-29-zph2.6-17483 & 150.31509 & 2.49175 & 2.4 \\
c17-29-zph2.6-34570 & 150.28142 & 2.48831 & 2.559 \\
c17-29-zph2.7-14804 & 150.31004 & 2.39676 & 2.566 \\
c17-29-zph2.8-14723 & 150.31305 & 2.45708 & 2.567 \\
c17-29-zsp2.3-01174 & 150.25618 & 2.38222 & 2.313 \\
c17-29-zsp2.6-01157 & 150.29317 & 2.45171 & 2.556 \\
c17-62-zph2.3-14818 & 150.34473 & 2.39424 & 2.279 \\
c17-62-zph2.4-14776 & 150.37248 & 2.41991 & 2.471 \\
c17-62-zph2.4-17503 & 150.33798 & 2.47542 & 2.403 \\
c17-62-zph2.5-14742 & 150.32265 & 2.44352 & 2.505 \\
c17-62-zph2.5-14763 & 150.32812 & 2.42992 & 2.555 \\
c17-62-zph2.6-14751 & 150.31796 & 2.43698 & 2.551 \\
c17-62-zph2.7-17517 & 150.3484 & 2.46721 & 2.587 \\
c17-62-zsp2.5-01159 & 150.35135 & 2.44302 & 2.452 \\
c17-62-zsp2.5-01168 & 150.31071 & 2.40391 & 2.496 \\
c17-62-zsp2.5-01502 & 150.3588 & 2.48178 & 2.471 \\
c17-62-zsp2.5-01512 & 150.35596 & 2.4634 & 2.471 \\
c17-61L-zph2.5-14882 & 150.34009 & 2.35964 & 2.452 \\
c17-61L-zph2.5-15010 & 150.32918 & 2.30061 & 2.317 \\
c17-61L-zph2.5-32231 & 150.32224 & 2.28384 & 2.586 \\
c17-61L-zph2.6-12224 & 150.31772 & 2.28078 & 2.534 \\
c17-61L-zph2.6-14940 & 150.36411 & 2.33619 & 2.494 \\
c17-61L-zph2.6-15018 & 150.37976 & 2.29622 & 2.496 \\
c17-61L-zph2.6-33462 & 150.37938 & 2.33715 & 2.498 \\
c17-61L-zph2.8-32238 & 150.38272 & 2.2858 & 2.658 \\
c17-61L-zsp2.5-01187 & 150.35432 & 2.35273 & 2.453 \\
c17-61L-zsp2.6-00767 & 150.31223 & 2.27923 & 2.578 \\
c17-61L-zsp2.7-01205 & 150.34872 & 2.32137 & 2.657 \\
c17-61L-zsp2.7-01212 & 150.37012 & 2.30588 & 2.655 \\
c17-60L-zph2.3-32330 & 150.35461 & 2.14912 & 2.272 \\
c17-60L-zph2.5-12291 & 150.35986 & 2.24675 & 2.455 \\
c17-60L-zph2.5-12400 & 150.32335 & 2.18807 & 2.502 \\
c17-60L-zph2.6-12359 & 150.36682 & 2.21295 & 2.483 \\
c17-60L-zph2.6-12375 & 150.32465 & 2.20695 & 2.49 \\
c17-60L-zph2.6-12432 & 150.36377 & 2.16507 & 2.576 \\
c17-60L-zph2.6-32321 & 150.36781 & 2.15833 & 2.614 \\
c17-60L-zph2.7-12334 & 150.38455 & 2.22249 & 2.728 \\
c17-60L-zph2.9-32309 & 150.31273 & 2.17346 & 2.67 \\
c17-60L-zsp2.3-00826 & 150.31908 & 2.16216 & 2.313 \\
c17-60L-zsp2.5-00794 & 150.31914 & 2.22503 & 2.493 \\
c17-60L-zsp2.6-00819 & 150.37843 & 2.17079 & 2.551 \\
c17-60L-zsp2.6-01719 & 150.31601 & 2.24457 & 2.583 \\
c17-60L-zsp2.7-00788 & 150.39108 & 2.24033 & 2.738 \\
c17-60L-zsp2.7-00802 & 150.34006 & 2.20841 & 2.729 \\
pc22L-zph2.3-17733 & 150.14896 & 2.4883 & 2.38 \\
pc22L-zph2.5-26344 & 150.10031 & 2.46256 & 2.477 \\
pc22L-zph2.8-33515 & 150.15773 & 2.4089 & 2.463 \\
p18-zph2.5-15320 & 150.00002 & 2.42489 & 2.63 \\
p18-zph2.6-15055 & 150.0636 & 2.42693 & 2.52 \\
p18-zph2.7-34724 & 150.05705 & 2.4823 & 2.392 \\
p18-zsp2.4-01220 & 150.06973 & 2.45253 & 2.423 \\
p15l-zph2.5-15385 & 149.94215 & 2.36591 & 2.477 \\
p15l-zph2.5-15435 & 149.95938 & 2.33549 & 2.506 \\
p15l-zph2.6-33610 & 149.99947 & 2.33633 & 2.583 \\
p15l-zsp2.6-01875 & 149.93594 & 2.29014 & 2.552 \\
p15l-zsp2.7-01354 & 149.94077 & 2.30644 & 2.681 \\
\end{longtable}
\end{center}

\begin{center}
\begin{longtable}{|l|l|l|}
\caption{\teff\ values and corresponding 1D errors for the \loz\ and \hiz\ redshift interval. The latter's values are appended to the former's and are separated by a row of dashes.} \label{tab:tau_vals} \\
\hline \multicolumn{1}{|c|}{\textbf{z}} & \multicolumn{1}{c|}{\textbf{\teff}} & \multicolumn{1}{c|}{\textbf{$\tau_{\sigma}$}} \\ \hline 
\endfirsthead
\multicolumn{3}{c}
{{\bfseries \tablename\ \thetable{}}} \\
\hline \multicolumn{1}{|c|}{\textbf{z}} & \multicolumn{1}{c|}{\textbf{\teff}} & \multicolumn{1}{c|}{\textbf{$\tau_{\sigma}$}} \\ \hline 
\endhead
\hline \multicolumn{3}{|r|}{{Continued on next page}} \\ \hline
\endfoot
\hline \hline
\endlastfoot
2.0285 & 0.2115 & 0.0926 \\
2.0315 & 0.0508 & 0.0761 \\
2.0345 & 0.0128 & 0.0736 \\
2.0376 & 0.0628 & 0.088 \\
2.0406 & 0.0167 & 0.0763 \\
2.0437 & 0.0372 & 0.0801 \\
2.0467 & 0.1107 & 0.0932 \\
2.0497 & 0.0136 & 0.084 \\
2.0528 & 0.1023 & 0.0828 \\
2.0742 & 0.1106 & 0.078 \\
2.0773 & 0.0145 & 0.0741 \\
2.0804 & 0.237 & 0.0857 \\
2.0835 & 0.2226 & 0.086 \\
2.0866 & 0.1933 & 0.0822 \\
2.0897 & 0.1386 & 0.0682 \\
2.0928 & 0.1717 & 0.0686 \\
2.0958 & 0.061 & 0.071 \\
2.0989 & 0.1111 & 0.0719 \\
2.102 & 0.2439 & 0.0763 \\
2.1052 & 0.2527 & 0.0732 \\
2.1083 & 0.2786 & 0.0772 \\
2.1114 & 0.3183 & 0.0822 \\
2.1145 & 0.1146 & 0.0651 \\
2.1176 & 0.0425 & 0.0682 \\
2.1207 & 0.0495 & 0.0684 \\
2.1238 & 0.1718 & 0.0663 \\
2.127 & 0.152 & 0.0657 \\
2.1301 & 0.2382 & 0.0723 \\
2.1332 & 0.0281 & 0.0759 \\
2.1364 & 0.3557 & 0.0854 \\
2.1395 & 0.3383 & 0.0831 \\
2.1426 & 0.3668 & 0.0716 \\
2.1458 & 0.3355 & 0.0665 \\
2.1489 & 0.279 & 0.0623 \\
2.1997 & 0.2233 & 0.058 \\
2.2029 & 0.2031 & 0.0545 \\
2.2061 & 0.3617 & 0.0625 \\
2.2094 & 0.3471 & 0.0649 \\
2.2126 & 0.1102 & 0.0527 \\
2.2158 & 0.0969 & 0.0503 \\
2.219 & 0.1706 & 0.0533 \\
2.2222 & 0.2714 & 0.0616 \\
2.2254 & 0.1884 & 0.0541 \\
2.2449 & 0.1062 & 0.0486 \\
2.2481 & 0.0974 & 0.0498 \\
2.2677 & 0.1961 & 0.0565 \\
2.2709 & 0.146 & 0.052 \\
2.2742 & 0.098 & 0.0474 \\
2.2775 & 0.1193 & 0.0455 \\
2.2808 & 0.1897 & 0.0492 \\
2.284 & 0.1481 & 0.0488 \\
2.2873 & 0.1862 & 0.0438 \\
2.2906 & 0.2515 & 0.0533 \\
2.2939 & 0.1222 & 0.046 \\
2.2972 & 0.2033 & 0.0458 \\
2.3005 & 0.1946 & 0.0572 \\
- & - & - \\
2.1566 & 0.3301 & 0.0692 \\
2.1597 & 0.3741 & 0.0739 \\
2.1629 & 0.196 & 0.0678 \\
2.1661 & 0.2386 & 0.0635 \\
2.1692 & 0.2195 & 0.0617 \\
2.1724 & 0.0898 & 0.0631 \\
2.1756 & 0.1067 & 0.0584 \\
2.1787 & 0.0725 & 0.0618 \\
2.1819 & 0.1125 & 0.0601 \\
2.1851 & 0.0345 & 0.0557 \\
2.2075 & 0.1165 & 0.0516 \\
2.2107 & 0.0854 & 0.061 \\
2.2139 & 0.1759 & 0.0558 \\
2.2171 & 0.1038 & 0.0536 \\
2.2204 & 0.2195 & 0.0607 \\
2.2236 & 0.1268 & 0.0482 \\
2.2268 & 0.11 & 0.0584 \\
2.23 & 0.2422 & 0.0587 \\
2.2333 & 0.1673 & 0.0534 \\
2.2365 & 0.2795 & 0.0576 \\
2.2397 & 0.1813 & 0.0534 \\
2.243 & 0.247 & 0.0515 \\
2.2462 & 0.1324 & 0.051 \\
2.2495 & 0.0812 & 0.0458 \\
2.2527 & 0.0554 & 0.0523 \\
2.256 & 0.0118 & 0.0503 \\
2.2592 & 0.0793 & 0.0481 \\
2.2625 & 0.0745 & 0.0489 \\
2.2658 & 0.128 & 0.057 \\
2.269 & -0.0031 & 0.0505 \\
2.2723 & 0.182 & 0.0436 \\
2.2756 & 0.2785 & 0.05 \\
2.2789 & 0.2218 & 0.0582 \\
2.2821 & 0.2094 & 0.0561 \\
2.2854 & 0.2975 & 0.0498 \\
2.3384 & 0.2053 & 0.0461 \\
2.3418 & 0.2232 & 0.0447 \\
2.3451 & 0.3334 & 0.0586 \\
2.3485 & 0.3247 & 0.0542 \\
2.3518 & 0.2601 & 0.0509 \\
2.3552 & 0.098 & 0.0411 \\
2.3585 & 0.2244 & 0.055 \\
2.3619 & 0.1617 & 0.05 \\
2.3652 & 0.1963 & 0.0481 \\
2.3855 & 0.0616 & 0.0463 \\
2.3889 & 0.0763 & 0.0377 \\
2.4093 & 0.1726 & 0.0505 \\
2.4127 & 0.204 & 0.0465 \\
2.4161 & 0.215 & 0.0362 \\
2.4195 & 0.1851 & 0.0385 \\
2.4229 & 0.1438 & 0.0376 \\
2.4264 & 0.1895 & 0.04 \\
2.4298 & 0.1142 & 0.0431 \\
2.4332 & 0.1336 & 0.04 \\
2.4367 & 0.1676 & 0.0468 \\
2.4401 & 0.1135 & 0.0408 \\
2.4436 & 0.1519 & 0.0428 \\
2.447 & 0.0025 & 0.0363 \\
\end{longtable}
\end{center}

\end{document}